\documentclass[twoside,11pt]{article}

\usepackage{melba}
\usepackage{float}
\usepackage{mwe} 

\usepackage{amsmath,amsfonts}

\melbaid{YYYY:NNN}  
\doi{https://doi.org/10.59275/j.melba.2023-AAAA}
\melbaauthors{Highton, Chong, Finestone, Beqiri, Schnabel and Bhatia}  
\volume{2}
\firstpageno{1}  
\melbayear{YYYY}  
\datesubmitted{m1/yyyy}  
\datepublished{m2/yyyy}  

\ShortHeadings{Robustness Testing Against CT Degradation}{Highton, Chong, Finestone, Beqiri, Schnabel and Bhatia}

\title{Robustness Testing of Black-Box Models Against CT Degradation Through Test-Time Augmentation}

\author{\firstname Jack \surname Highton \email \\  
	\addr School of Biomedical Engineering and Imaging Sciences, King's College London, London, UK; \\ 
 Aival Ltd., London, UK
	\AND
	\name Quok Zong Chong \email \\
	\addr Aival Ltd., London, UK
        \AND
        \name Samuel Finestone \email \\
	\addr Aival Ltd., London, UK
        \AND
        \name Arian Beqiri \email \\
        \addr School of Biomedical Engineering and Imaging Sciences, King's College London, London, UK \\
	\addr Aival Ltd., London, UK
        \AND
        \name Julia A. Schnabel \email \\
	\addr School of Biomedical Engineering and Imaging Sciences, King's College London, London, UK; \\
 School of Computation, Information and Technology, Technical University of Munich, Munich, Germany; \\
 Helmholtz Munich, Munich, Germany
        \AND
        \name Kanwal K. Bhatia \email kanwal@aival.io \\
	\addr Aival Ltd., London, UK
}

\begin{document}
\nolinenumbers

\maketitle

\begin{abstract}
	Deep learning models for medical image segmentation and object detection are becoming increasingly available as clinical products. However, details are rarely provided about the training data, thus models may unexpectedly fail when cases differ from those in the training distribution. An approach allowing potential users to independently test a model’s robustness, treating it as a `black-box' and using only a few cases from their own site, is key for adoption. To address this, a method to test the robustness of these models against CT image quality variation is presented. In this work we present this framework by demonstrating that given the same training data, the model architecture and data pre-processing greatly affect the robustness of several frequently used segmentation and object detection methods to simulated CT imaging artifacts and degradation. Our framework also addresses the concern about the sustainability of deep learning models in clinical use, by considering future shifts in image quality due to scanner deterioration or imaging protocol changes which are not reflected in a limited local test dataset.
\end{abstract}


\section{Introduction}

Segmentation and object detection are important medical imaging analysis tasks in clinical and research settings. Segmentation is a fundamental analysis step because it provides boundary localization and volume quantification of anatomical and pathological structures which may be key for diagnosis and treatment planning ~\citep{giorgio2013clinical} ~\citep{cabral1993interactive}. Automated object detection can enhance radiological reporting by highlighting pathological features ~\citep{choi2022deep}, or can guide further medical image analysis to focus on important features of the image ~\citep{jin2022object}. However, a lack of clinicians' trust in deep learning based applications can undermine adoption, which can be remedied with scientific robustness testing of an application using data from the clinicians own center ~\citep{ahmad2021reviewing}. Within this field, robustness refers to a model's ability to maintain performance when encountering data which differs from the training dataset, due to a shift in demographics, acquisition protocol, or acquisition artifacts ~\citep{galati2022accuracy}.

X-ray Computed Tomography (CT) is a frequently used medical imaging modality with many applications, including radiotherapy planning, tumor diagnosis, angiography and trauma analysis ~\citep{liguori2015emerging}. This work presents a method to facilitate model independent robustness testing using a limited amount of CT data. 

Deep neural networks based models have recently become the foremost approach for automating many segmentation and object detection tasks and can achieve expert human performance in some applications ~\citep{kooi2017large}. However, models have various vulnerabilities which can make them less robust than human visual assessment ~\citep{geirhos2018generalisation}. This issue becomes particularly acute as deep learning systems are increasingly distributed as commercial products, where restricted information about the model architecture and training data render the application an effective ‘black-box’ for the end user, making their robustness unclear and undermining trust. 

Robustness testing of a deep learning model should consider the types of input degradation it would encounter during real-world use, where degradation is the process by which the quality of an image is diminished or compromised. Although it is well known that models are susceptible to adversarial examples ~\citep{szegedy2013intriguing}, these cases are created by an agent deliberately attempting to fool the system which is unlikely in a medical imaging context. Also, although common degradations affecting natural image quality, e.g. noise, contrast alterations, and blurring, have been shown to affect models' performance ~\citep{geirhos2018generalisation}, the degradations seen in medical images have a fundamentally different nature. The key concern for robustness in a medical imaging application is degradation caused by the acquisition process. This may be caused by the acquisition protocol changing, such as the resolution or CT exposure parameters. There may also be image artifacts, defined as image features not present in reality but which appear due to unintentional acquisition phenomena, such as the patient moving during the scan or the effect of metal implants.

For a model trained on data with consistent acquisition parameters and without notable artifacts, these degradations can result in a discrepancy between the characteristics of training and test data, called a distribution shift ~\citep{quinonero2008dataset}. Test data affected by this discrepancy is a form of out-of-distribution (OOD) data, which in medical imaging context may be frequently encountered and which can undermine the robustness of deep learning applications. Although there are many approaches to retrospectively mitigate CT artifacts, including simulations to train models to remove them ~\citep{van2022generating}, there is little understanding of how specific parameter changes and artifacts in CT imaging can create a distribution shift which affects the performance of models and therefore their safety in a medical context.

\begin{figure}[H]
		\centering
		\includegraphics[width=1.0\linewidth]{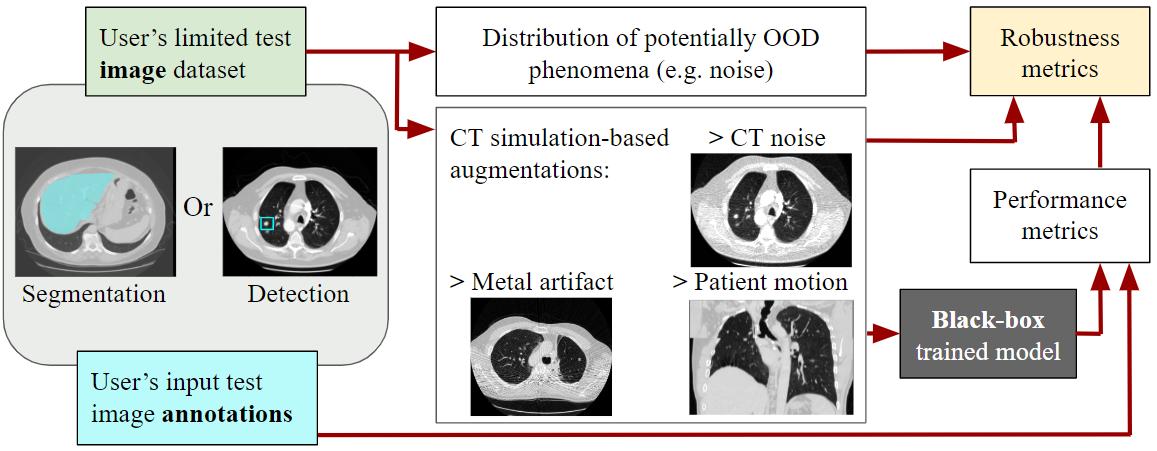}
	\caption{An overview of the proposed framework for robustness testing. The potential user of a black-box model wishes to test its robustness to CT image degradation using a small locally acquired test dataset, annotated with segmentation maps or object boxes outlined by an expert. The performance of the model is evaluated after various simulation-based augmentations are applied to the input image (center). Then the evaluation results, alongside information from the test dataset noise distribution, are used to calculate robustness metrics for the model.}
    \label{fig:over}
    \end{figure}

We address this by developing a framework to systematically assess robustness of black-box models for segmentation and object detection with CT images (see Figure \ref{fig:over}). This would allow a user with a limited test dataset acquired from their own center, with annotations representing the optimal output for those cases according to an expert, to test a segmentation or object detection model against OOD cases not present in their dataset but which are likely to occur during future use. Our suite of tests include increased CT noise, artifacts due to metal implants, and patient motion during the scan. Our framework allows direct comparison of the robustness of multiple ‘black-box’ models designed for the same task. This aims to aid confidence in the decision to adopt deep learning based methods in clinical practice, as well as trust in the continuing robustness of these models to future changes in image quality.  

\section{Related work}

\subsection{Identifying OOD images with knowledge of the model and training data}

The most direct method to test a models's robustness against OOD data is to evaluate its performance on a dataset which is known to be OOD compared to the model's training data. In medical imaging, this approach has been used to test the accuracy of OOD detection algorithms designed to automatically identify cases which the model is not expected to be robust to ~\citep{vasiliuk2023redesigning} ~\citep{vasiliuk2023limitations} ~\citep{nguyen2023out}. OOD test sets were compiled by selecting datasets with a known distribution shift relative to the training dataset, such as different acquisition protocols or underlying patient conditions. However, this is dependent on knowledge about the training data, which is not available unless provided by the developers.

Methods do exist to explore if input data are OOD when the training dataset is not known. If access to a model's architecture and trained weights is available, methods like Mahalanobis distance ~\citep{anthony2023use} and generalized ODIN ~\citep{vasiliuk2023limitations} can be used to predict if an input medical image is OOD. In other situations the models weights may not be accessible, such as when it has been integrated into a product or if model obfuscation methods have been applied ~\citep{zhou2023modelobfuscator}. In this case, the output softmax confidence score can be used to predict if an input is OOD ~\citep{hendrycks2016baseline} ~\citep{liu2020energy}. However, OOD detection methods based on confidence output have limited accuracy in natural image applications ~\citep{liu2020energy}, which becomes significantly worse for medical imaging applications in segmentation ~\citep{vasiliuk2023limitations}. Also, obtaining the softmax output may depend on having access to the final layer of the neural network architecture before a threshold is applied.

\subsection{Testing robustness against OOD images from benchmark datasets}

A sufficiently diverse benchmark dataset can provide a fully model-agnostic method to test a black-box model's robustness against OOD data \citep{boone2023rood}. An ideal benchmark dataset would contain a range of input images with corresponding annotations and where any potentially OOD properties of each input image is labeled. This allows evaluation of the accuracy of the final output against the annotations, summarizing how much the model's performance degrades when it encounters each given type of OOD data. An example benchmark dataset contained cardiac Magnetic Resonance Imaging (MRI) images labeled according to the imaging center, scanner type, and disease condition, which formed part of a challenge to create generalizable segmentation models ~\citep{campello2021multi}. However, the report from the challenge organizers did not discuss what properties of images in certain groups led to reduced performance from the submitted models ~\citep{campello2021multi}. MOOD 2020 is another example benchmark dataset for brain MRI and abdominal CT~\citep{zimmerer2022mood}. Some images were labeled by humans as ODD due to anomalous pathology, but most were created by augmenting images from the in-distribution set. The categories of augmentations included removal of slices, blurring, global and local deformations, and randomly inserting patches of noise and sections of other images. Some have questioned whether these augmentations realistically depict anomalous medical images in the real world~\citep{li2023self}, suggesting the augmented image patches should be formed by blurring multiple extracted image features together. However, the patch based augmentations still would not reflect realistic OOD cases which result from the acquisition process. 

\subsection{Testing robustness against images with acquisition based augmentations}

Augmentations to create potentially OOD images can instead be designed to recreate acquisition phenomena known to cause anomalous cases in the real world. For example, a patient may enter the scanner in an unusual position due to a spinal injury ~\citep{yang2023imaging}. This could be reflected by image translation and rotation augmentations, which are already frequently included as default training augmentations \citep{goceri2023medical}. It has been found that MRI hardware and software updates can change image properties enough to affect segmentation methods and thus distort brain tissue volume results in longitudinal studies \citep{medawar2021estimating} \citep{potvin2019measurement}. Changing the scanner or its parameters may alter the image resolution or field of view, which can be recreated by the common training augmentations of rescaling and cropping \citep{goceri2023medical}.

A shift in the image noise distribution is another effect of changing scanning parameters, such as when the CT radiation dose changes by altering the tube current or slice scan time \citep{goldman2007principles}. Even though adding Gaussian distributed noise is a common training-time augmentation technique \citep{goceri2023medical}, in the medical imaging domain actual noise has a non-Gaussian distribution associated with the acquisition modality. Speckle, Rician and Poisson distributed noise have been suggested as training-time augmentations for ultrasound \citep{singla2022speckle}, MRI \citep{boone2023rood}, and planar X-Ray \citep{khalifa2022comprehensive} images respectively. The noise in CT scans has a complicated spatial structure, augmentation of which requires either a physics-based simulation of the CT acquisition process \citep{won2014realistic} or the use of a generative model \citep{liu2022learning}. A physics-based simulation of CT noise has been used to augment the training data for model's to classify lung image patches for the presence of lung nodules \citep{omigbodun2019effects}, which did not lead to a performance improvement for test data of various noise levels compared to Gaussian noise augmentation. However, this study only considered one specific task, and did not use the physics based model to assess the robustness of different model architectures to CT noise.

Patient motion during a scan can cause image artifacts resulting in image degradation. A simulation of MRI motion artifacts has been been used to augment the training data of a segmentation model, which improved its robustness against real world artifacts \citep{shaw2020k}. 

A recent study \citep{boone2023rood} provided a benchmark dataset to test the robustness of models for MRI segmentation against OOD cases, created using a series of physics-based augmentations. This included noise, illumination field, resolution changes, spatial transformations and motion artifacts. The authors also defined a degradation metric to quantify a model's robustness against a certain augmentation, which takes into account the mean or standard deviation of the segmentation performance metrics (e.g. Dice \citep{dice1945measures}) after a range of augmentation severity is applied to a test set. Since they found that the variance of the segmentation performance metrics increases with more extreme augmentations, a similar degradation metric using the variance was suggested. Both approaches allow direct comparison of the robustness of two models when tested with the same augmented dataset, by comparing the degradation value. 

\subsection{Testing robustness against OOD images with limited datasets}

It is crucial for the end user to be able to independently test cases from their own clinical center to support their decision to adopt deep learning based applications for medical imaging analysis in a clinical context \citep{jacobson2021clinical}, because they must be appropriately trained for the targeted population demographics \citep{vayena2018machine}, disease presentation, and acquisition system \citep{prior2020open}. Testing data has to be manually labeled by human experts, resulting in high cost and limited size of testing datasets \citep{koh2022artificial}, which is often exacerbated at a local level \citep{shaikhina2017handling}. 

If the test dataset is limited in size, the statistical significance of summative performance metrics (e.g. mean Dice) may be undermined and it may lack rarer types of OOD images which may be encountered in the future. Methods have been developed to expand small medical imaging datasets with synthetic data produced by generative adversarial networks (GANs), mostly to train models for classification tasks such as COVID-19 detection~\citep{waheed2020covidgan}. GANs have also been used to expand a training dataset for 2D retina image segmentation~\citep{noguchi2020bone}. Since this approach also requires the generation of accurate corresponding synthetic ground truth segmentation masks, the segmentation model being trained has to be integrated into the data synthesizing algorithm, rendering this an inappropriate way to generate test data to independently assess black-box models. Non-deep learning approaches for expanding datasets mix features in the images to create new images. A cropping and patching method has been proposed for both images and segmentation mask annotations ~\citep{noguchi2020bone}, however the images produced contain anatomy which is concatenated in a clearly unrealistic way. Laplacian blending has been proposed as a more realistic alternative for classification datasets ~\citep{sanaat2022robust}, but this method would pose problems for generating valid annotations in the transition regions for both segmentation and object detection applications.   

The above methods to expand limited datasets can generate cases which are potentially OOD due to new combinations of image features. However, the patch based methods combine image features in an unrealistic way, while GAN based methods create ground truths which are not independent of the model being tested. Furthermore, these methods cannot add OOD cases resulting from acquisition artifacts/degradation not already present in the dataset. This motivates our CT simulation based framework, which is independent of the models being tested and allows them to be treated as a black-box. The augmentations generated by the simulation can also add common artifacts/degradations not initially present, or ones with greater severity, to the limited test dataset.

\subsection{Contributions}

To address these issues with existing approaches to independently test the robustness of black-box models with limited datasets, for the first time to our knowledge, we develop a benchmarking platform to create datasets for evaluating the robustness of models to corruptions and artifacts in CT. Analogous to previous work \citep{boone2023rood} which presented a system for robustness and out-of-distribution testing specifically to brain MRI segmentation models (ROOD-MRI), here we expand and further generalize this concept to more generic segmentation and localization road-testing tasks in CT imaging. We make the following scientific contributions to this field:

\begin{itemize}

  \item We propose a physics-based CT simulation for modeling associated acquisition anomalies: noise variation, metal implant artifacts, and patient motion. Our simulation approach only uses parameters obtained from analysis of the CT images themselves, allowing use when details of the CT acquisition are not available.

  \item Using test-time data augmentations from the simulation, we demonstrate that different deep learning model architectures have weaker robustness to different CT artifacts. 
  
  \item We investigate established models, designed for object detection (retinaNet) and for segmentation (U-Net, nnUnet). We also convert the tested segmentation models into object detection applications, to compare their robustness with retinaNet.

  \item Inspired by ROOD-MRI \citep{boone2023rood} which uses fixed weighting of the contribution of each augmentation severity level to calculate the summative noise degradation metric, here we expand on that formula by using an empirical distribution of noise in a given dataset to calibrate the weighting of noise augmentation severity levels. 

\end{itemize}

The remainder of our paper is structured as follows. In Section \ref{materials} we outline the datasets we used to train models. The established model architectures that we trained to apply the robustness testing framework to are briefly described: 3D-UNet \citep{cciccek20163d}, nnUnet \citep{isensee2021nnu}, and retinaNet \citep{lin2017focal}. Then, we describe the metrics used to evaluate their performance at segmentation or object detection using test sub-datasets. Section \ref{meth for aug} describes how our CT simulation approach is based only on parameters extracted from the test CT images themselves, allowing general use to generate the augmentations for robustness testing. In Section \ref{results}, we show these applied to generate a summary robustness metric calibrated with the dataset. Then our results show how these methods demonstrate the strengths and weakness of different model architectures. 

\section{Materials and metrics} \label{materials}
    \subsection{Datasets} \label{datasets}
        \subsubsection{LUNA 16}
        
LUNA 16 is a lung nodule object detection challenge dataset \citep{luna16web}. We included it in this study because lung nodule annotation is crucial for lung cancer diagnosis, but manual labeling is time-consuming and error prone, so deep learning applications are of significant interest \citep{ren2020unsupervised}.

The dataset includes CT scans with a maximum slice thickness of 2.5 mm from the publicly available LIDC/IDRI database \citep{armato2011lung}. Annotations were collected by 4 experienced radiologists, and the LUNA 16 annotations consist of all nodules larger than 3 mm accepted by at least 3 out of 4 radiologists. The annotations were presented as ground truths for object detection, where all nodules were labeled as having the same class while their position and size were defined by 3D bounding box co-ordinates. We used 570 images for training models while 30 were left aside for testing.

The training images were pre-processed by resampling via bilinear interpolation to (0.703, 0.703, 1.25 mm) voxel sizes and the intensity was windowed to -1024 to 300 Hounsfield Units (HU). This pre-processing was applied to the test data after the application of CT simulation based augmentations described in Section \ref{meth for aug}.
        
    \subsubsection{Segmentation decathlon - liver task}

As part of the The Medical Segmentation Decathlon challenge \citep{antonelli2022medical}, the liver CT dataset consists of contrast-enhanced CT images from patients with primary cancers and metastatic liver disease. We included this dataset due to the importance of robust organ segmentation for treatment planning, such as radiotherapy \citep{yu2022multi}. 

The dataset contained target segmentation maps outlining the liver. It was acquired in the IRCAD Hôpitaux Universitaires, France and contained a subset of patients from the 2017 Liver Tumor Segmentation challenge \citep{bilic2023liver}. We used 570 images for training models and 30 were left aside for testing. 

The training images were pre-processed by resampling via bilinear interpolation to (1.5, 1.5, 1.5 mm), but the intensity was not windowed by default because the effect of training models with and without windowing was compared during robustness testing. This pre-processing was applied to the test data after the application of CT simulation based augmentations described in Section \ref{meth for aug}.

    \subsection{Segmentation model architectures and training} \label{models} 
    
        \subsubsection{3D-UNet}

The implementation of 3D-UNet \citep{cciccek20163d} segmentation model provided within the MONAI library \citep{cardoso2022monai} was utilized. 3D-UNet was chosen because it has frequently been applied as a baseline architecture to compare the performance of newly developed models against during the past five years \citep{li2023automatic} \citep{li2023automatic} \citep{bui2019multi}. We trained this architecture with three pre-processing approaches for the training data, to test the effect they have on robustness:

    \begin{itemize}

    \item No image augmentation, apart from random sampling of patches for training with size (96,96,48) for the liver decathlon dataset and (192,192,80) for LUNA 16.
  
    \item Image augmentation using the protocol used for nnUnet shown in Table \ref{table_nnUnet_props}, followed by the random patch selection described above. 

    \item Image intensity windowing in the range -60 to 160 HU, followed by the random patch selection described above. This option was only used for the liver segmentation dataset because the window corresponds to the range of intensity of the liver and surrounding volume, while the intensity of lung nodules in the LUNA 16 dataset is very variable.
    
    \end{itemize}
        
        \subsubsection{nnUNet}

    An integration of the nnUnet \citep{isensee2021nnu} segmentation protocol within the MONAI library \citep{monai_nnunetv2} was used. This trains an ensemble of UNet derived models with parameter optimization based on the data and hardware before selecting the best performing combination, as outlined in Table \ref{table_nnUnet_props}. nnUnet was chosen because it is considered a state-of-the-art segmentation protocol, after achieving the best performance across multiple medical imaging segmentation applications in the The Medical Segmentation Decathlon challenge \citep{antonelli2022medical}.    

    \begin{table}[H] 
        \centering
        \caption{Comparison of the properties of the nnUNet and base 3D-UNet used here. \\
        *The largest patch size possible with Batch Size 2 given the available memory. \\
        **The best pair of the models, ensembled by averaging softmax probabilities, chosen by cross validation of the training data.}
        \begin{tabular}{ccc}
            \textbf{Dataset} & \textbf{nnUNet} & \textbf{3D-UNet}  \\
            \hline
            Training Augmentation & Rescale x0.9 or 1.2, 15\% & None \\
                                  & Gaussian Noise std 0.01, 15\% &   \\
                                  & Gaussian Smooth x0.5-1.15, 15\% &   \\
                                  & Intensity -0.3 or 0.3, 15\% & \\
                                  & Flip (each axis), 50\% & \\
            \\
            Learning Rate & `poly' schedule (initial 0.01) & 0.001 \\
            Training Epochs & 1000 & 600 \\
            Batch Size & 2 & 1  \\
            Loss Function & Dice + Cross Entropy & Dice  \\
            Intensity Normalization & clipped: 0.5 to 99.5\% & clipped: -57 to 164 HU  \\
            In-plane Image Resampling & In plane 3rd order spline &  Linear interpolation \\
            Slice Image Resampling & Nearest Neighbor & Linear interpolation  \\
            In-plane Annotation Resampling & Linear interpolation & Linear interpolation  \\
            Slice Annotation Resampling & Nearest Neighbor & Linear interpolation  \\
            Patch Size & Maximum for Batch Size 2* & 96x96x96 \\
            Low-resolution Patch Size & 25\% of median image size & N/A  \\
            \\
            Ensemble Selection Options & 3D-UNet & N/A \\
                               & 2D-UNet &   \\
                               & Low-resolution UNet cascade &   \\
                               & Ensemble of two** &   \\
            \hline
        \end{tabular}
        \label{table_nnUnet_props}
    \end{table}

    \subsection{Object detection model architectures and training}

            \subsubsection{retinaNet}

    We utilized an implementation of the retinaNet \citep{lin2017focal} architecture provided within the MONAI library. retinaNet is composed of a backbone network (ResNet in this implementation) with several downsampling layers followed by several upsampling layers. Each of the upsampling layers is sampled by two convolutional sub-networks to predict the object class and bounding box co-ordinates, giving independent box predictions considering different resolution scales which allows accurate identification of both large and small objects. This is appropriate for the LUNA 16 dataset, which contains annotated nodules varying widely in diameter from 3.0 mm to 28.3 mm. retinaNet also uses a focal loss function for classification, cross entropy weighted by uncertainty so difficult to classify examples are more heavily weighted, but there was only one class of nodule considered so this feature was not relevant.  

            \subsubsection{Conversion of segmentation models to object detection models} \label{conversion}

    Object detection models are evaluated here using the mean Average Precision (mAP), while the segmentation models are evaluated using the Dice score. Therefore, in order to allow direct comparison, we added a post-processing step to the segmentation models applied to the LUNA 16 dataset to create bounding boxes for the nodules like an object detection model, which can be evaluated using mAP. To do this, contiguous volumes of voxels labeled as `nodule' within the predicted segmentation maps with a volume greater than 14.14 mm\textsuperscript{3} were selected. This is the volume of a sphere with a 3mm diameter, the minimum size of annotated nodules in the LUNA 16 dataset. For each of these contiguous volumes, the minimum enclosing box with edges parallel to the orthogonal image axes was calculated and treated as a box produced by an object detection model with that `nodule' class at 100\% confidence.

        \subsection{Evaluation metrics} \label{metrics}
        \subsubsection{Segmentation evaluation}
        
The Dice coefficient \citep{dice1945measures} (DSC) is a \textit{de facto} standard \citep{maier2024metrics} used to quantify the accuracy of a segmentation method \citep{zijdenbos1994morphometric}:
\begin{equation}
DSC = \frac{{2 \times |A \cap B|}}{{|A| + |B|}}
\end{equation}
where $A$ and $B$ represent the set of ground truth and the inferred segmentation voxels, and $|A|$ and $|B|$ are the amount of voxels in those sets. DSC ranges from 0 to 1, where 1 represents perfect overlap between the ground truth and inference. The mean DSC across a test dataset is used to evaluate the accuracy of a segmentation model. The consistency of the model may be impacted when OOD images are encountered, reflected by an increase in the standard deviation of DSC values \citep{boone2023rood}.

    \subsubsection{Object detection evaluation}

The mean Average Precision (mAP) is an evaluation metric for object detection methods used across several benchmark challenges \citep{padilla2020survey}, which summarizes the precision-recall trade-off dictated by confidence levels of the predicted bounding boxes.

mAP is the mean of the Average Precision (AP) for each class, where AP is the area under a precision recall curve that has been preprocessed to remove zig-zag behavior \citep{padilla2021comparative}. To calculate the precision and recall at each threshold confidence value, each bounding box predicted by the model with a confidence exceeding the threshold is compared with the ground truth boxes using the metric Intersection over Union (IoU) \citep{padilla2021comparative}:
\begin{equation}
IoU = \frac{{|A \cap B|}}{{|A \cup B|}}
\end{equation}
If the IoU between a predicted box and a ground truth box of the corresponding class exceeds a specified threshold, the predicted box is a true positive. If the IoU does not meet the threshold, another predicted box has already met the threshold, or the predicted class is wrong, then the predicted box is a false positive. Any ground truth boxes which does not have a corresponding true positive are counted as false negatives. The precision and recall can then be calculated for each threshold confidence value. mAP can be calculated for a single test image or across a test set. It should be noted that a `black-box' object detection model may have an intrinsically fixed confidence threshold, in which case AP reduces to precision multiplied by recall.

mAP is often calculated with a range of IoU thresholds, with the mean result taken \citep{padilla2020survey}. This approach is taken here, because the models being compared may perform best at different IoU thresholds. Furthermore, the IoU may be heavily affected by how tightly the image is bounded by the ground truth box, so a range of IoU thresholds helps capture this variability. This is particularly the case for annotations of lung nodules, which vary greatly in size, have boundaries which are difficult for a clinician to precisely discern \citep{larici2017lung}, and which have a varying heterogeneous shape compared to an enclosing cuboid.

\section{Methods for augmentation of CT data} \label{meth for aug}
    \subsection{CT simulation}

A custom physics based simulation of CT was used to generate CT specific augmentations. This was based on the Astra toolbox \citep{van2016fast}, which creates the system geometry then numerically simulates the stages of the acquisition process.

The pre-processing of the input images to be augmented follows a previously outlined method \citep{won2014realistic}. First, the image CT values (Hounsfield Units) are converted to attenuation X-ray attenuation values $\mu$ (cm\textsuperscript{-1}):
\begin{equation}
\mu = \frac{{\text{{CT number}}}}{{1000}} \times \mu_{\text{{water}}} + \mu_{\text{{water}}}
\end{equation}
where $\mu_{\text{{water}}}$ is the linear attenuation coefficient of water which is 0.18 cm\textsuperscript{-1} at 120 kVp \citep{boedeker2007application}. Then, total variation based denoising \citep{rubin1992nonlinenr} was used to remove noise from the original acquisition from the attenuation image. 

    \begin{figure}[H]
		\centering
		\includegraphics[width=0.8\linewidth]{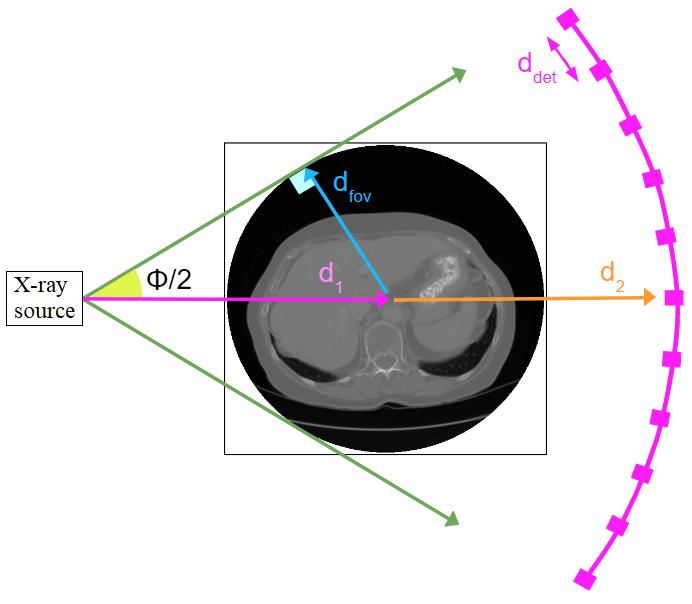}
	\caption{The geometry parameters of the simulated CT system: $\phi$ beam angle, d\textsubscript{1} source to image center distance, d\textsubscript{2} image center to array distance, d\textsubscript{det} distance between adjacent detectors, and n\textsubscript{det} is the number of detectors.}
    \label{fig:geom}
    \end{figure}

\begin{figure}[H]
		\centering
		\includegraphics[width=1.0\linewidth]{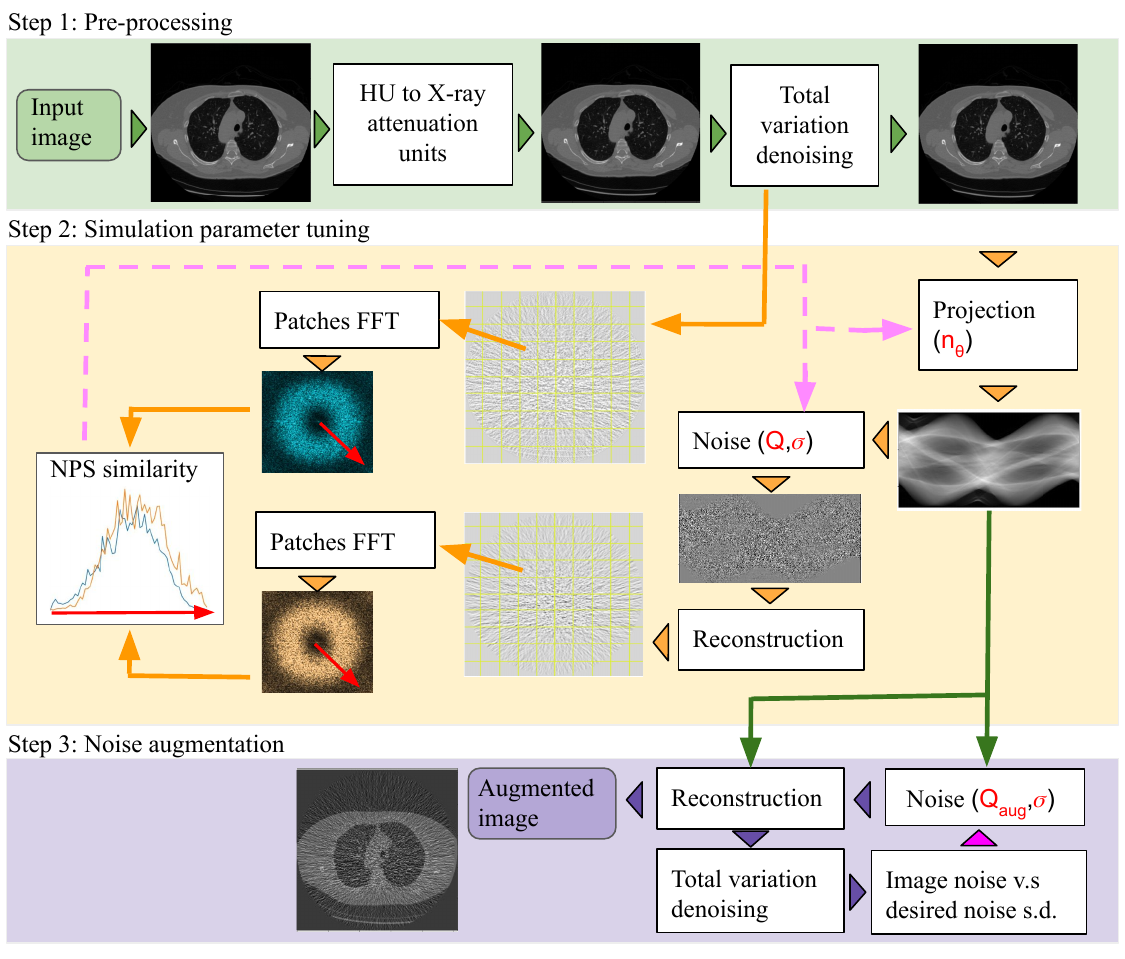}
	\caption{Outline of the CT noise simulation tuning process. Step 1: The image intensity is converted from HU to attenuation units before denoising and noise extraction via the total variation method. Step 2: the simulation parameters Q\textsubscript{0}, $\sigma$, and n\textsubscript{$\theta$} are selected to maximize the similarity in the radial noise power spectrum (NPS) patchwise between the noise generated by the simulation and the noise extracted from the original image, as described in Table \ref{fig:search}. Step 3: To increase the level of noise, the incident flux Q\textsubscript{0} is decreased until a desired augmented image noise level is reached, as described in Section \ref{noise_gen}.}
    \label{fig:tune}
    \end{figure}

An outline of the geometry of the CT scanner system is required to simulate the acquisition process, but as this is often not available from image metadata, we use the following procedure to allow general CT augmentation. The geometric parameters required by the Astra toolbox are outlined in Figure \ref{fig:geom}: source to image center distance d\textsubscript{1}, image center to array distance d\textsubscript{2}, number of x-ray detectors n\textsubscript{det}, and distance between adjacent detectors d\textsubscript{det}. We assumed the beam angle $\phi$ to be 60\textdegree  as is typical of modern scanners \citep{peyrin2021ct}, although a specific value is sometimes available in the metadata. We also assumed n\textsubscript{det} to equal 1500, which is consistent with modern scanners \citep{hermena2021ct} and higher values did not change the reconstructed image. The other geometric parameters can be calculated as follows.

d\textsubscript{1} can be obtained using $\phi$ and the diameter of the circular field of view d\textsubscript{fov}. d\textsubscript{fov} can measured from the input CT image by identifying the edge of the circular field of view, which is achieved by sampling a line of voxels from the corner of the central slice of the image to the image center, until an intensity value other then the uniform value outside the field of view is encountered. 

\begin{equation}
d\textsubscript{1} = \frac{d_{fov}}{sin(\phi / 2)}
\end{equation}
d\textsubscript{2} can be assumed to be similar (approximated as equal) to d\textsubscript{1} because both the X-ray source and detectors are attached to the same rotating gantry, but the exact value will not affect the simulation if the air is assumed to have negligible attenuation and the values of n\textsubscript{det} and d\textsubscript{det} correspond to the projected size of the detector array arc given d1 and d2. 
\begin{equation}
d_{det} = \frac{2 \:\: (d_1 + d_2) \:\: tan(\phi / 2)}{n_{det}} 
\end{equation}
Another parameter required by the acquisition simulation is the number of sampled angles n\textsubscript{$\theta$}, but this was empirically tuned alongside the noise parameters in the process outlined in Figure \ref{fig:tune}.

The sinogram generated from the projection stage of the simulation was used to reconstruct a CT image using Astra's filtered back projection function with the default Ramachandran-Lakshminarayan filter \citep{ramachandran1971three}.

Once the projection stage of the simulation has generated a sinogram for each slice, augmented noise could be added. This can be modeled as a combination of Poisson noise due to a restricted number of photons reaching the detector, and Gaussian electronic noise \citep{won2014realistic}. As our augmentation approach is designed to be general, including for when specific information about the beam properties is not available, we empirically tune the incident flux of photons on each each detector point with no X-ray attenuation Q\textsubscript{0}. This, in combination with the attenuation of the X-ray beam by the scanned object, determines the actual incident number of photons and thus Poisson noise, and Gaussian noise parameter $\sigma$. The sinogram after application of the noise augmentation $S_{ns}$ is
\begin{equation}
S_{ns} = S_{0} \, + \, P( \, Q\textsubscript{0} \, exp(-S_{0}) \, ) \, + \, G(\sigma)
\end{equation}
where $S_{0}$ is the initial sinogram, $P(x)$ is a random sample from the Poisson distribution with expected value x, $G(\sigma)$ is a random sample from a Gaussian distribution with a zero mean and standard deviation $\sigma$.

The noise model has three parameters which require empirical tuning using the input dataset (see Table \ref{fig:search}). We did this by comparing the CT noise generated by the model with the noise extracted from the input test images by total variation denoising (see Figure \ref{fig:tune}). This utilized each test image by breaking the central slice into a 10x10 grid, calculating the radial noise power spectrum (NPS) in each cell \citep{dolly2016practical}, calculating the mean of the sum of square differences between the NPS curves from the data and model, mSSE\textsubscript{NPS}. The parameters were tuned to minimize the mSSE\textsubscript{NPS}, which represents dissimilarity in the CT noise accounting for its spacial variation. The tuning process was a two stage coarse-to-fine grid search, using the search parameters in Table \ref{fig:search}.

    \begin{table}[H] 
        \centering
        \caption{The unitless simulation parameters tuned to recreate the CT noise extracted from the test dataset, by minimizing the mean noise power spectrum dissimilarity (mSSE\textsubscript{NPS}). The fine  grid search parameters were found by applying the multiplications or additions shown to the optimum parameter found in the coarse search stage.}
        \begin{tabular}{cccc}
            \textbf{Parameter} & \textbf{Description} & \textbf{Coarse Search} & \textbf{Fine Search} \\
            \hline
            Q\textsubscript{0} & Incident photon flux & $10^4$,$10^5$,$10^6$,$10^7$ & x0.5, x0.75, x1.0, x2.5, x5 \\
            $\sigma$ & s.d. of Gaussian noise & 0, 0.1, 1, 10 & x0.5, x0.75, x1.0, x2.5, x5 \\
            n\textsubscript{$\theta$} & Number of sampled angles & 720,1440,2160,2880 & -360, +0, +360 \\
        \end{tabular}
        \label{fig:search}
    \end{table}

    \subsection{Artifact generation}
        \subsubsection{Noise} \label{noise_gen}

The noise in the image produced by the CT simulation can be increased by reducing the value of Q\textsubscript{0}, modeling a lower X-ray dose, until the desired standard deviation (s.d.) of the simulated noise field is obtained. This search is done by rounding the optimized Q\textsubscript{0} to the nearest order of magnitude (10\textsuperscript{$m$}, where $m$ is an integer) and decreasing $m$, then fine-tuning using multiples of: [0.5, 0.75, 1.0, 1.25, 1.5], to obtain the closest s.d. value.

\begin{figure}[H]
		\centering
		\includegraphics[width=0.8\linewidth]{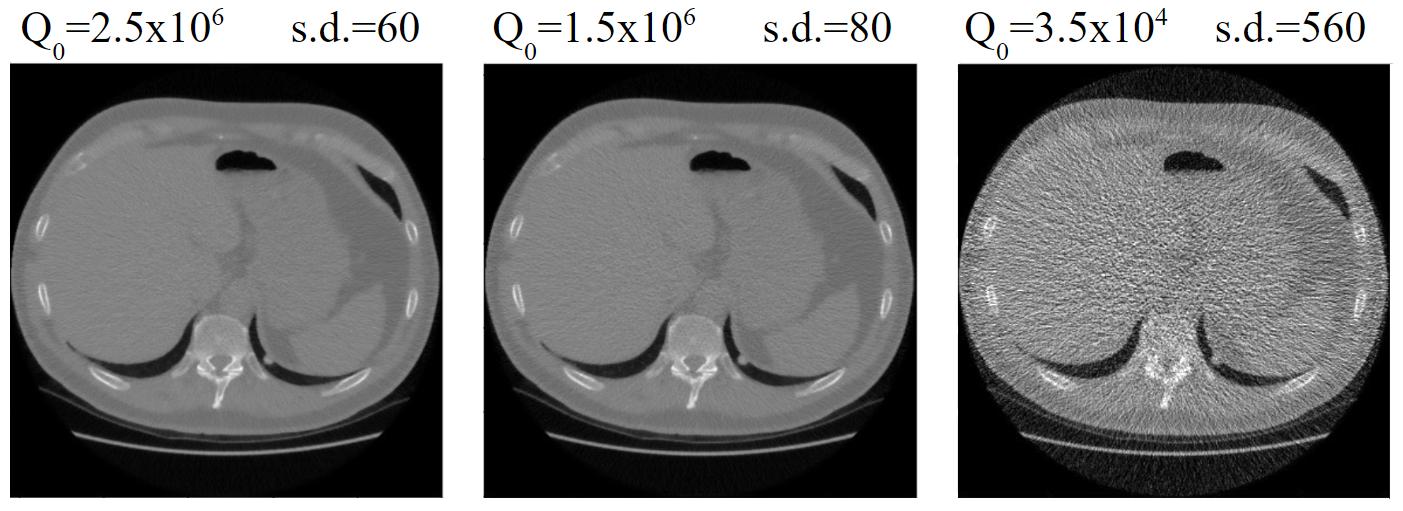}
	\caption{The CT noise augmentation is performed by reducing the the simulation parameter Q\textsubscript{0}, to the value resulting in the output images having an increased noise component with a given standard deviation (s.d.).} 
    \label{figure:noise91}
    \end{figure}

\begin{figure}[H]
		\centering
		\includegraphics[width=0.8\linewidth]{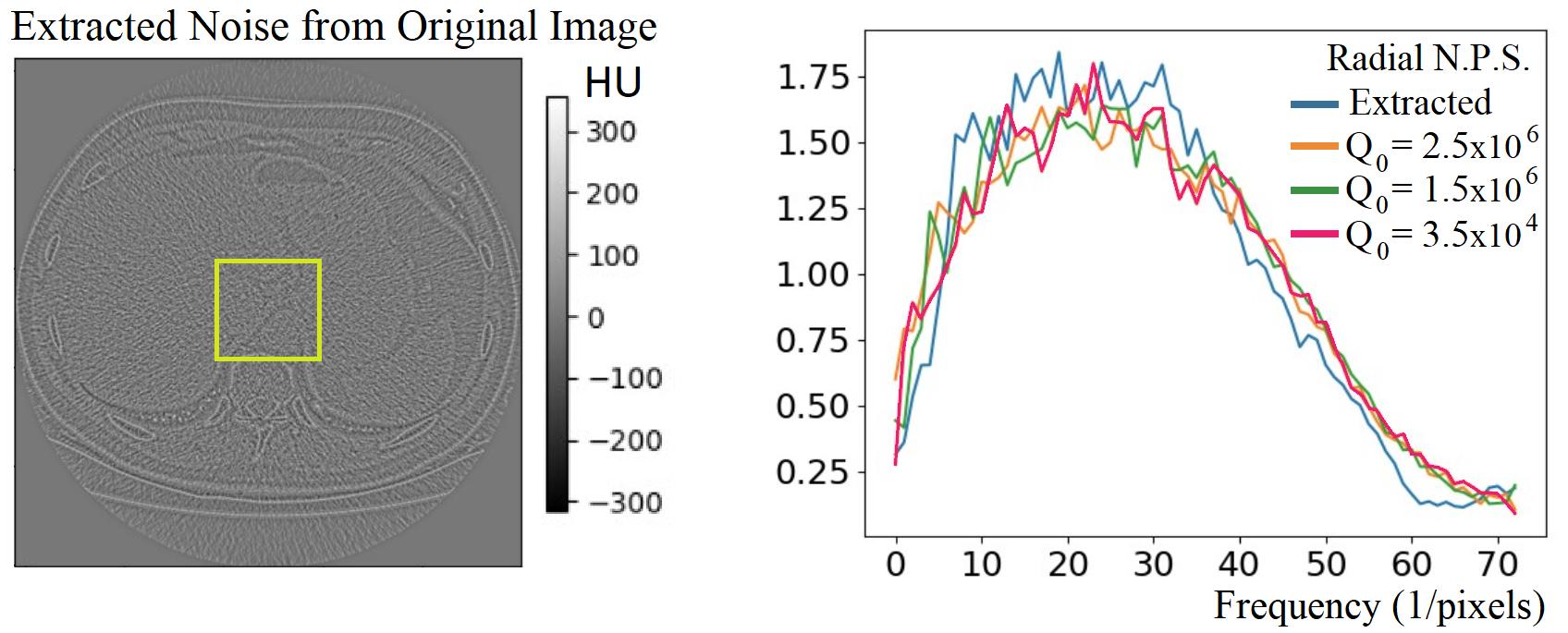}
	\caption{The radial noise power spectrum (NPS) is sampled from the noise extracted from the original image within a central patch (left) for comparison with the same patch of the simulated noise fields. The normalized NPS's are shown to have similar profiles (right).}
    \label{figure:nps91}
    \end{figure}

As shown by the simulated images in Figure \ref{figure:noise91} and the noise power spectra in Figure \ref{figure:nps91}, this method of increasing the noise level does not significantly change the spatial distribution or texture of the noise field. Therefore, the strength of this augmentation can be quantified by the noise s.d., even though the augmentation is caused by altering the physical parameter Q\textsubscript{0}. 

        \subsubsection{Metal implants}

Metallic implants can cause streaking artifacts in CT images. In order to simulate this phenomenon, we augment the input images slices by inserting a cylindrical region with a radiodensity of 20,000 Hounsfield units, corresponding to pure steel \citep{bolliger2009differentiation}. These implants were centered at the highest intensity point in the spine and orientated axially with a length spanning the axial range of the annotated objects or segmentation plus an extra 15mm in both directions. This arrangement was chosen because the image slices which contain part of the metal implant are affected by the characteristic streaking artifact, so in the augmented images generated by the the CT simulation, this artifact would be present in the slices containing the annotated objects or segmentation and also the region superior and inferior to them. Therefore, the effect of the metal implant artifact surrounds the annotated objects or segmentation.

\begin{figure}[H]
		\centering
		\includegraphics[width=1.0\linewidth]{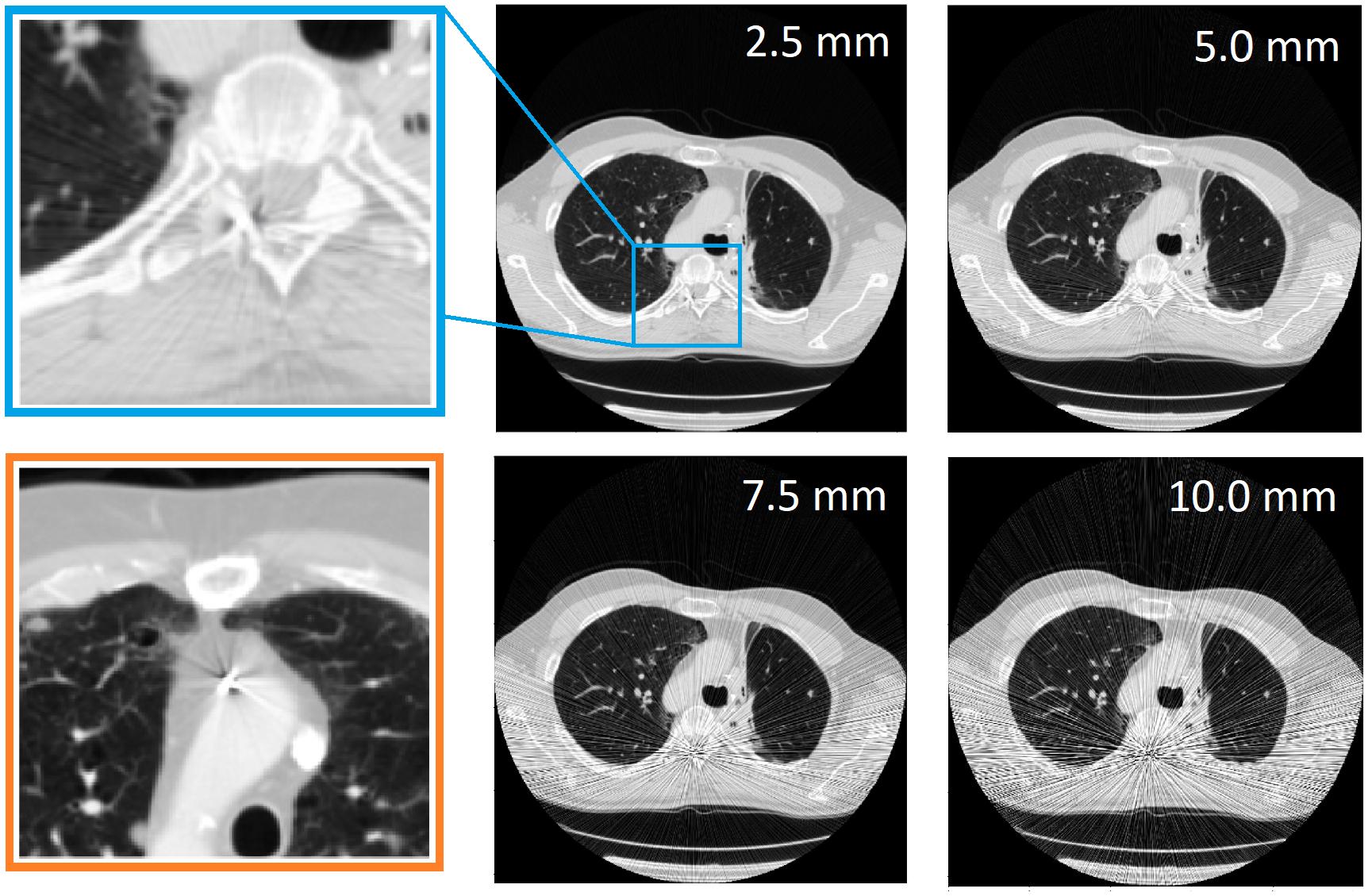}
	\caption{The simulation of a metallic implant as a form of CT data augmentation causes a streaking artifacts with a severity depending on the implant radius (white numbers). An example of the simulated artifact due to a 2.5mm radius implant (blue outline) is visually similar to an example artifact found in the LUNA 16 training dataset (orange outline).}
    \label{figure:streak6707}
    \end{figure}

We scaled the augmentation strength by increasing the radius of the implants. The output of the CT simulation contained the characteristic streaking artifacts after the input images were augmented with a metal implant. For small implants the artifacts were visually similar to examples in the LUNA 16 dataset, as shown in Figure \ref{figure:streak6707}. Simulated implants with a larger radius caused more extensive streaking, which is visually similar to reported observations of artifacts due to spinal implants \citep{barrett2004artifacts}. However, comparably extensive streaking was not seen in the LUNA 16 dataset.

\clearpage
        
        \subsubsection{Motion}

We model patient motion as a rigid change in pose by tilting during the sequential acquisition of slices, by applying an axial rotation to all slices above or below a certain point and thus producing a discontinuity in the CT image. We consider two types of motion augmentation, which differ due to how their severity is scaled:
\begin{itemize}

\item{Motion magnitude (mag): The augmentation strength is increased by increasing the amount of rotational motion in degrees, while the location of the slice discontinuity is constant at 10mm inferior to lowest point of the annotation.} 
\item{Motion proximity (prx): The augmentation strength is increased by decreasing the distance of the motion discontinuity to the to the lowest point of the annotation, while the amount of rotation is kept constant at 10\textdegree.}

\end{itemize}

\section{Experiments and results} \label{results}

    \subsection{Dataset analysis} \label{noise anal}

\begin{figure}[H]
		\centering
		\includegraphics[width=1.0\linewidth]{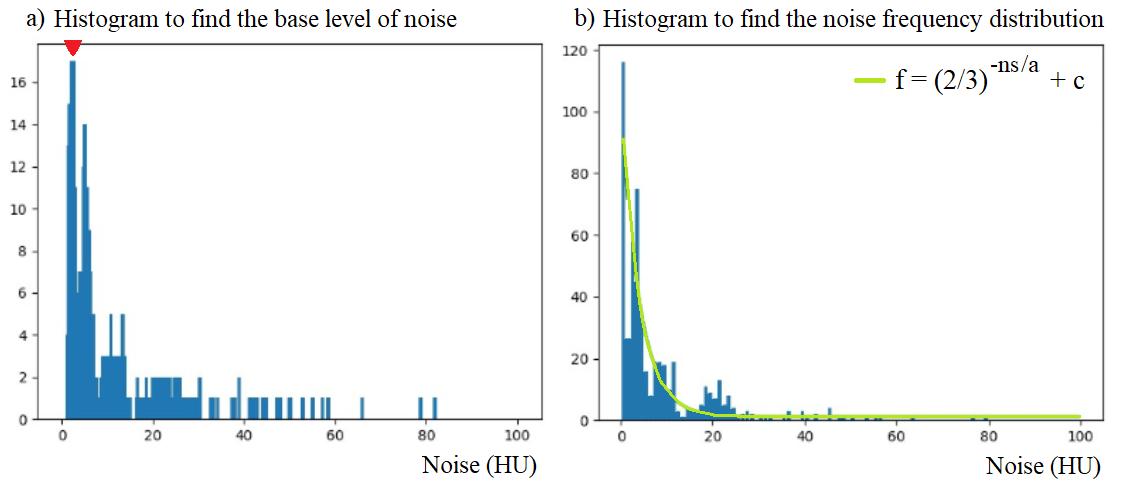}
	\caption{Measurements of the noise in the original LUNA 16 dataset allow empirical scaling of the degradation weights. The standard deviation of the noise extracted by the total variation method is measured for each case. a) We use a histogram to identify the modal level of noise as a base level (found to be 4 HU). b) We use the cases with a noise equal or higher than the base level to produce a second histogram, to which the distribution function (green) is fitted. Both histograms use a bin width of 1 HU.}
    \label{noise_hist}
    \end{figure}

Figure \ref{noise_hist} shows the measurement of the noise distribution in the LUNA 16 dataset. This is used to find the set of weights w\textsubscript{s}, replacing the coefficient w\textsubscript{s} = (2/3)\textsuperscript{s} in equation (\ref{deg_eq}) \citep{boone2023rood}, where s is the level of augmentation severity.

\begin{equation}
Deg_{TM} = \frac{1}{\sum_{s=1}^5 \omega_s} \sum_{s=1}^5 \omega_s (mM_{clean} - mM_{T,s})  \:\:, \:\:\: \text{with:} \:\:\:\:\:  \omega_s = (2/3)^s   
\label{deg_eq}
\end{equation}

We obtained the weights (0.221,0.044,0.006,0,0,0) corresponding to the noise augmentation levels of (10,20,50,100,200,350,500) by dividing the frequency distribution at those points by its value at the base level of noise (see Figure \ref{noise_hist}b).

    \subsection{Noise simulation} \label{noise res}

\begin{figure}[H]
		\centering
		\includegraphics[width=0.9\linewidth]{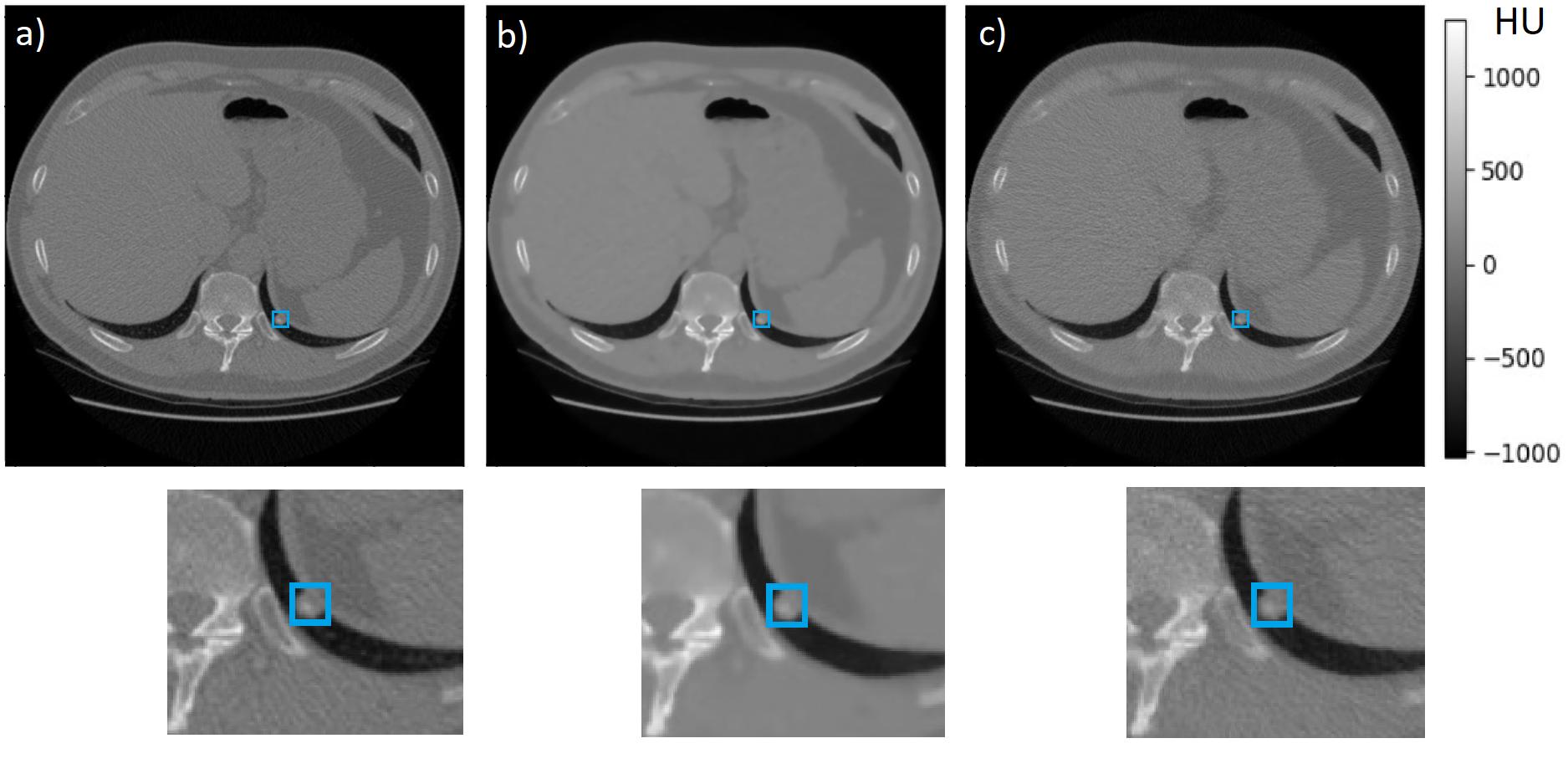}
	\caption{a) A slice from the LUNA 16 dataset, with a nodule labeled with a blue box. b) The slice with after denoising using the total variation method. c) the simulation is shown to recreate visually similar CT noise.}
    \label{fig:luna91regen}
    \end{figure}

An example from the LUNA 16 dataset with high noise is shown in Figure \ref{fig:luna91regen}. The CT noise is removed by total variation based denoising \citep{rubin1992nonlinenr} and recreated by the simulation, showing good visual similarity with the original image noise. In this case, the tuned simulation values are: Q\textsubscript{0} = $2.5 \times 10^6$, $\sigma$ = $0$, and n\textsubscript{$\theta$} = 2160. 

    \subsection{Model degradation results overview}

An overview of the performance and degradation scores shows that nnUnet was overall the most robust segmentation model (see Table \ref{tab_dce}), while retinaNet was overall the most robust object detection model (see Table \ref{tab_map}). These summary results are analyzed closely in the following subsections.

    \begin{table}[H] 
        \centering
        \caption{Degradation results for the segmentation models, where a lower score means better robustness ($\downarrow$). The DSC is measured between the annotation segmentation map and the model prediction, where a higher Base DSC means better model performance before augmentation ($\uparrow$). 3D-UNet (aug) is the 3D-UNet architecture with the same set of training augmentations used as nnUnet. 3D-UNet (win) is the 3D-UNet architecture with pre-processing of the input image by intensity windowing with the range -60 to 160 HU.}
        \begin{tabular}{cc|c|cccccc}
        \textbf{Dataset} & \textbf{Model} & \textbf{Base DSC $\uparrow$} & \multicolumn{4}{c}{\textbf{Degradation metric $\downarrow$}} \\
        & & & Noise & Metal & Motion mag & Motion prx\\
            \hline
            Liver Seg. & 3D-UNet      & 0.808 $\pm$ 0.09 & 0.123 & 0.111 &  0.002 & 0.001 \\
            ~       & 3D-UNet (aug)   & 0.862 $\pm$ 0.09 & 0.128 & 0.106 &  0.003 & 0.000 \\
            ~       & 3D-UNet (win)   & 0.908 $\pm$ 0.10 & 0.217 & 0.155 &  0.000 & 0.001 \\
            ~       & nnUNet          & \textbf{0.962} $\pm$ 0.10 & \textbf{0.061} & \textbf{0.070} & \textbf{0.000} & \textbf{0.000} \\
            \hline
            LUNA 16 & 3D-UNet       & 0.751 $\pm$ 0.07 & 0.050 & 0.167 & 0.037 & \textbf{0.027} \\
            ~       & 3D-UNet (aug) & 0.700 $\pm$ 0.08 & 0.046 & \textbf{0.143} & \textbf{0.023} & 0.041\\
            ~       & nnUNet        & \textbf{0.830} $\pm$ 0.07 & \textbf{0.031} & 0.166 & 0.038 & 0.051\\
        \label{tab_dce}
        \end{tabular}
    \end{table}

    \begin{table}[H] 
        \centering
        \caption{Degradation results for 3D Object Detection models, where a lower score means better robustness ($\downarrow$). For each image, mAP is the mean result for the IoU thresholds (0.01, 0.15... 0.9), where a higher base mAP means better performance before augmentation ($\uparrow$). 3D-UNet (aug) is the 3D-UNet architecture with the same set of training augmentations used as nnUnet. The nnUNet and 3D-UNet segmentation models are converted to detection models producing object bounding boxes, see Section \ref{conversion}.}
        \begin{tabular}{cc|c|cccccc}
        \textbf{Dataset} & \textbf{Model} & \textbf{Base mAP $\uparrow$} & \multicolumn{4}{c}{\textbf{Degradation metric $\downarrow$}} \\
        & & & Noise & Metal & Motion mag & Motion prx\\
        \hline
        LUNA 16 & RetinaNet     & \textbf{0.709} $\pm$ 0.20 & \textbf{0.001} & \textbf{0.228} & \textbf{0.003} & 0.023  \\
        ~       & 3D-UNet       & 0.549 $\pm$ 0.17 & 0.067 & 0.243 &  0.037 & 0.033 \\
        ~       & 3D-UNet (aug) & 0.591 $\pm$ 0.19 & 0.068 & 0.231 &  0.023 & 0.052 \\
        ~       & nnUNet        & 0.660 $\pm$ 0.17 & 0.085 & 0.264 &  0.038 & \textbf{0.015} \\
        \label{tab_map}    
        \end{tabular}
    \end{table}

    \subsection{CT noise augmentation}

    \begin{figure}[H]
		\centering
		\includegraphics[width=1.0\linewidth]{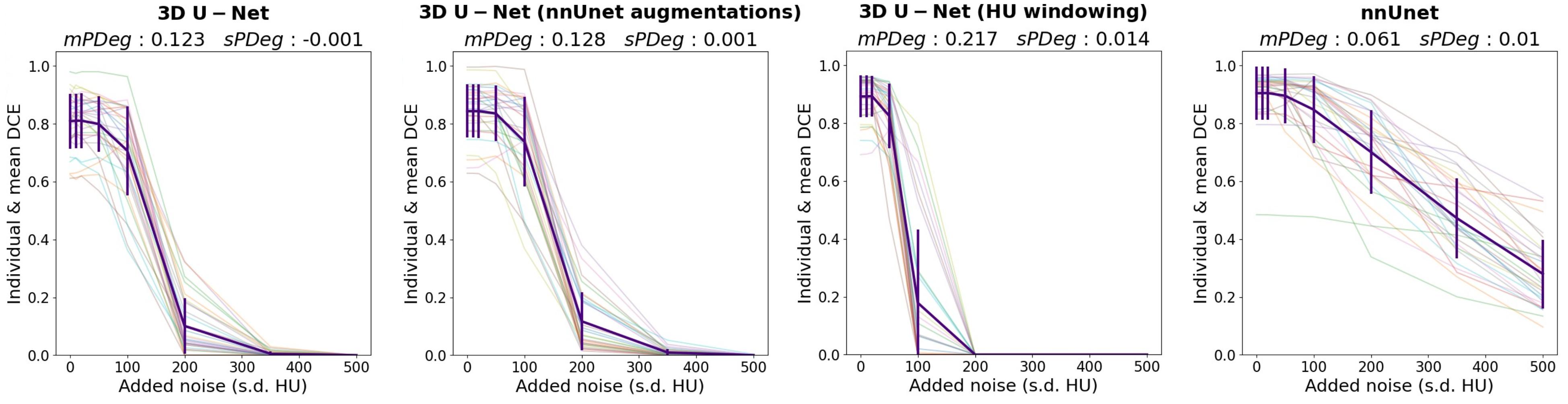}
	\caption{The Dice (DSC) score results against the level of CT noise added by augmentation, for the segmentation models each tested with 30 cases from the liver segmentation dataset. The means with standard deviation error bars are shown in purple. nnUnet demonstrates higher robustness than the 3D-Unet models, shown by the shallower drop off in DSC with increasing noise and lower degradation metrics.}
    \label{liver noise 3}
    \end{figure}

As shown in the `Noise' column of Table \ref{tab_dce} and plotted in Figure \ref{liver noise 3}, applying noise augmentation to the liver segmentation decathlon dataset shows that the nnUnet is more robust to increased CT noise than the 3D-UNet, as the mean DSC degradation (mDDeg) is lower. The robustness of the 3D-UNet increases when the nnUnet suite of training augmentation transforms was applied (see Table \ref{table_nnUnet_props}). Therefore, part of the increase in robustness is due to the training time augmentations, and part due to the model ensemble selection protocol and parameter tuning utilized by nnUnet. 

\begin{figure}[H]
		\centering
		\includegraphics[width=1.0\linewidth]{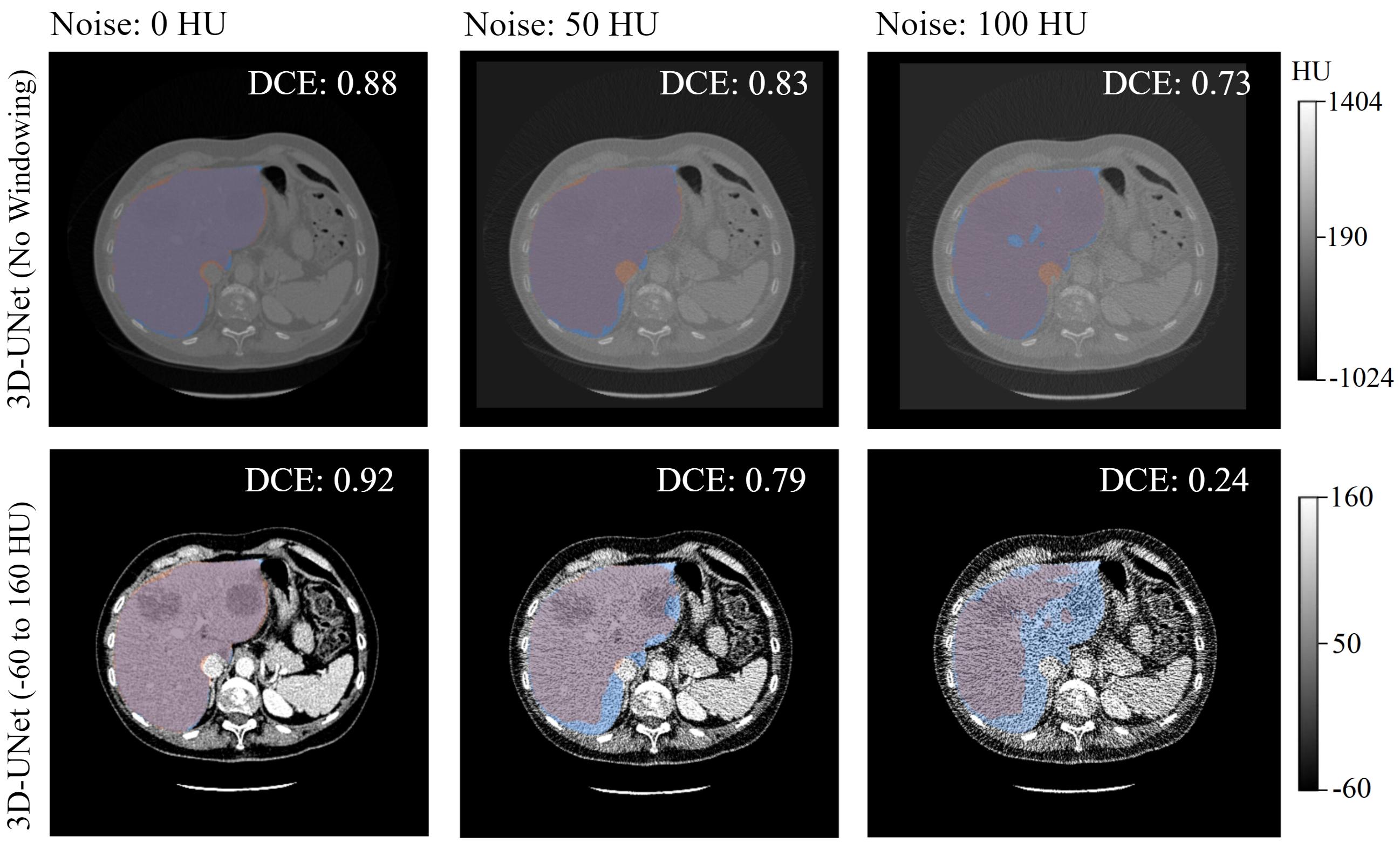}
	\caption{Top: The output segmentation for the 3D-UNet model (without nnUnet based augmentation). Bottom: The same model trained with intensity windowing as a -60 to 160 HU pre-processing step shown below. The annotation segmentation map is blue, the predicted segmentation map is orange, and the overlap is purple. The windowing results in the effect of the noise is greater relative to the range of intensity values, resulting in the DSC reducing faster as the simulated noise level increases. Note the windowed images would not be available to a user if the windowing was integrated into the model, but we are able to reveal this as the model developer.}
    \label{liver wind}
    \end{figure}

Adding contrast windowing as part of the training and test prepossessing for 3D-UNet (HU windowing) resulted in greatly reduced robustness. This can be expected, because the model is trained to handle a smaller range of intensity values (-60 to 160 HU) and sufficient CT noise results in values being beyond this range. Although the effect of windowing on the input image shown in Figure \ref{liver wind} would not be visible to the user if this pre-processing step is intrinsic to a black-box model, the difference in robustness demonstrated by our framework would be. 

As shown in Table \ref{tab_map}, testing with the LUNA 16 test dataset shows that the retinaNet object detection model is very robust to increased CT noise, with no positive degradation, in contrast to any of the object detection models derived from segmentation models. When the robustness of the segmentation models were tested with the DSC metric, the nnUnet is more robust that the 3D-UNet. Adding the nnUnet suite of training augmentation transforms did not improve the robustness of 3D-UNet at this task.

A notable response of the 3D-UNet and nnUnet models to increasing CT noise is the great increase in DSC and mAP standard deviation. This is due to the metrics improving for some of the cases, which is because some false positive nodule segmentation or box generation were avoided when the nodule-like image features were obscured by noise. However, for the whole dataset this is outweighed by less true positive outputs with increased noise augmentation. This effect shows the importance of the standard deviation based degradation metrics (sMDeg and sPDeg) in complementing the metric based degradation metrics (mMDeg and mPDeg).

    \subsection{Metal implant augmentation}

\begin{figure}[H]
		\centering
		\includegraphics[width=0.9\linewidth]{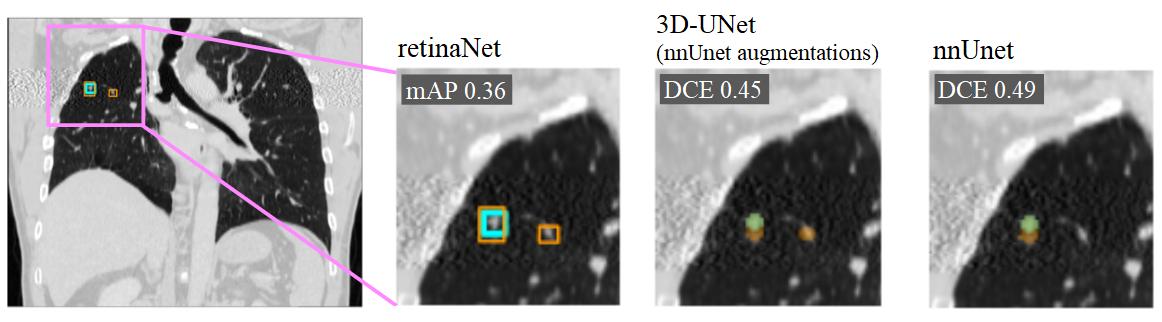}
	\caption{A coronal view of the streak artifacts produced by a simulated 7.5 mm radius cylindrical implant which extends 15mm above and below the annotated nodules, as well as the model outputs after this augmentation on a case from LUNA 16. A higher mean Average precision (mAP) means the output object detection boxes (orange) more accurately reflect the annotations (blue). A higher Dice score (DSC) means the output segmentation maps (orange) have a greater overlapping volume (green) with the annotation maps (blue).}
    \label{cyl_cor}
    \end{figure}

\begin{figure}[H]
		\centering
		\includegraphics[width=1.0\linewidth]{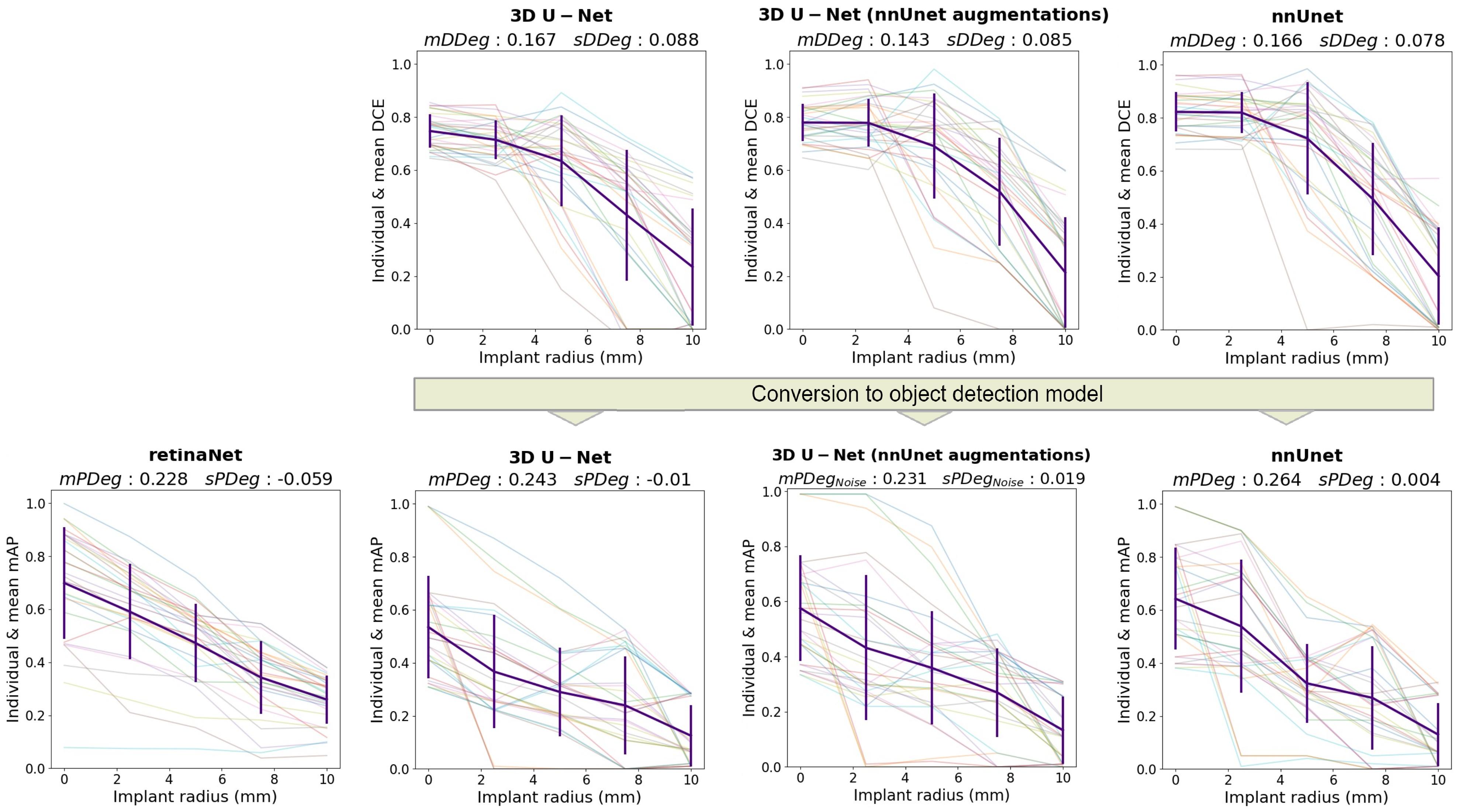}
	\caption{Top: The Dice (DSC) score results against simulated implant radius for the tested segmentation models. Bottom: mean Average precision (mAP) results against simulated implant radius for the tested object detection models, including segmentation models converted to generate bounding boxes. These are shown for the 30 test cases from the LUNA 16 dataset, with a cylindrical spinal implant added which induces a simulated streak artifact. The means with standard deviation error bars are shown in purple. The models degrade at a similar rate as the implant radius is increased, but retinaNet is more robust than the converted detection models.}
    \label{luna cyl implant 7}
    \end{figure}

As shown in the `Metal' column of Table \ref{tab_dce}, applying a simulated cylindrical implant to the liver segmentation decathlon dataset shows that the nnUnet is more robust to streak artifacts than the 3D-UNet. Like during the noise augmentation testing, the robustness of the 3D-UNet increased when the nnUnet suite of training augmentation transforms is applied. 
Table \ref{tab_dce}, Table \ref{tab_map} and Figure \ref{luna cyl implant 7} also show this when the cylindrical implant augmentation was applied to the LUNA 16 test dataset and the segmentation models were tested. Figure \ref{luna cyl implant 7} also shows that retinaNet is more robust to streak artifacts than the object detection models derived from segmentation models. Figure \ref{cyl_cor} shows the streak artifacts across a band of image slices caused by a simulated cylindrical spinal metal implant in a case from LUNA 16, as well as how this impacts the models' output.  

    \subsection{Motion augmentation} \label{mot res}

\begin{figure}[H]
		\centering
		\includegraphics[width=1.0\linewidth]{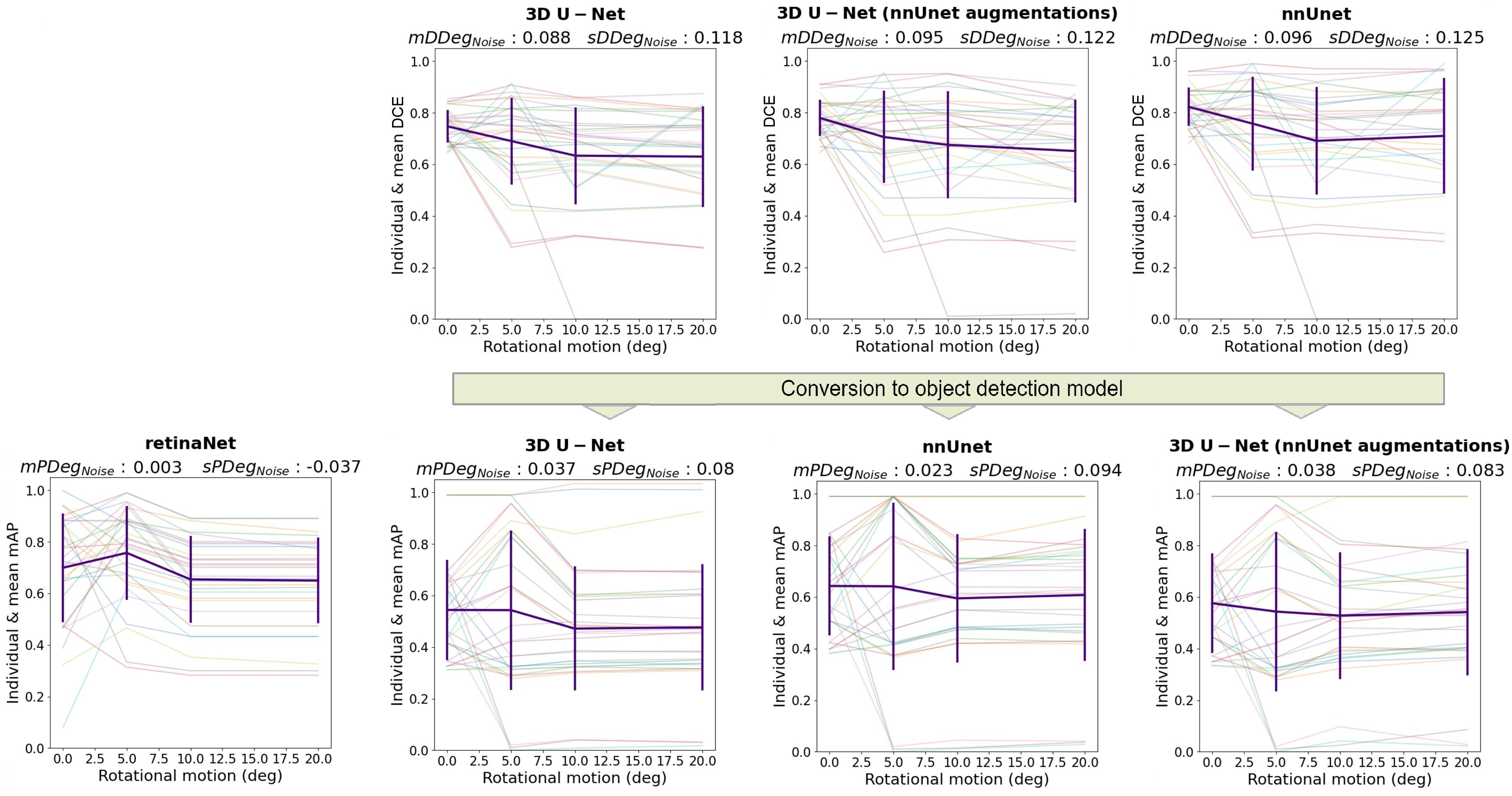}
	\caption{Top: The Dice (DSC) score results against the amount of rotational motion for the tested segmentation models. Bottom: mean Average precision (mAP) results against the amount of rotational motion for the tested object detection models, including segmentation models converted to generate bounding boxes. These are shown for the 30 test cases from the LUNA 16 dataset, with the discontinuity resulting from the single tilt motion located 5mm inferior to the lowest nodule containing slice according to the annotation. The means with standard deviation error bars are shown in purple. The models did not significantly degrade as the rotation increased beyond 10 degrees.}
    \label{luna 5 20 rot}
    \end{figure}

The `Motion mag' column of Tables \ref{tab_dce} and \ref{tab_map} show that all of the models tested were robust to rotational motion augmentation when the mean metric based degradation were considered. The models' performance do not significantly degrade as the magnitude of the rotational motion is increased beyond 10 degrees - see Figure \ref{luna 5 20 rot}. However, the standard deviation based degradation (sDDeg and sPDeg) shows a significant effect for the LUNA 16 test dataset, as the rotational discontinuity 5mm from the lowest nodule caused both increases and decreases in the segmentation DSC. 


\begin{figure}[H]
		\centering
		\includegraphics[width=1.0\linewidth]{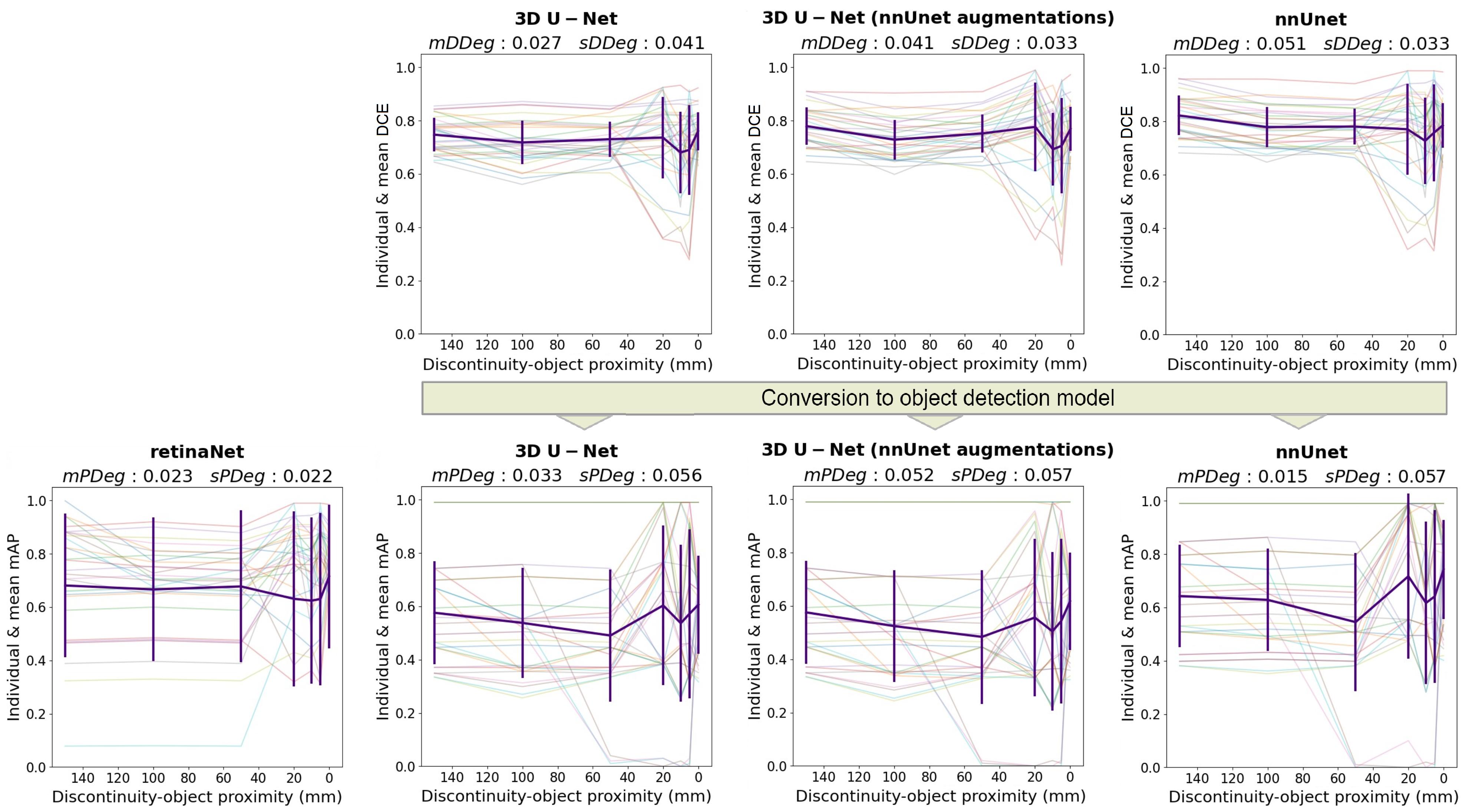}
	\caption{Top: The Dice (DSC) score results against proximity of the motion discontinuity from the lowest point containing a nodule, for the tested segmentation models. Bottom: mean Average Precision (mAP) results against proximity of the motion discontinuity from the lowest point containing a nodule, for the tested object detection models including segmentation models converted to generate bounding boxes. These are shown for each of the 30 test cases from the LUNA 16 dataset with 10 degrees of rotational motion causing a discontinuity at varying proximity to the lower edge of the lowest nodule according to the annotations. The means with standard deviation error bars are shown in purple. The models' performance degrade and then improve as the discontinuity approaches the annotation.}
    \label{luna vary rot od}
    \end{figure}

The reason for this is shown by the plots in Figure \ref{luna vary rot od}. When the discontinuity is shifted very close to the annotated object, the discontinuity sometimes improves the segmentation or detection result as the discontinuity matches the annotated object boundary in the axial direction. This effect is also seen for object detection models.

\begin{figure}[H]
		\centering
		\includegraphics[width=1.0\linewidth]{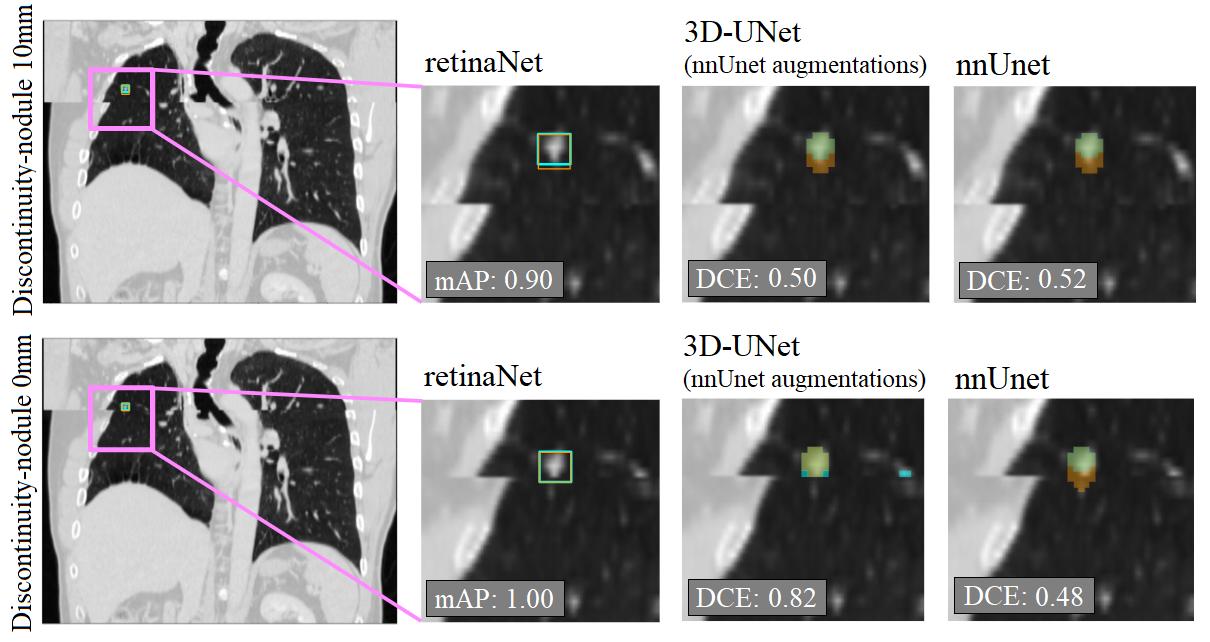}
	\caption{Discontinuity-nodule 10mm (top row): A coronal view of the image discontinuity produced by a 10 degree patient rotation 10mm below the annotated nodule, as well as the model outputs after this augmentation on a case from LUNA 16. Discontinuity-nodule 0mm (bottom row): The corresponding outcome with the discontinuity immediately below the annotated nodule, showing the 3D-UNet and RetinaNet performance improving due to the sharp boundary coinciding with the bottom of the annotated nodule.}
    \label{rot_dist_comparison}
    \end{figure}

 An example of this effect is shown in Figure \ref{rot_dist_comparison}, where the model outputs improve when the discontinuity is adjacent to the nodule. This suggests a random motion discontinuity location should be chosen for robustness testing, rather than one based on annotation, to avoid giving the tested models implicit information which they would not have in real world use.

\section{Discussion and conclusion}

In this work, we present a framework combining augmentations and evaluation metrics for evaluating the robustness of segmentation and object detection models with OOD cases produced by simulated protocol changes and artifacts in CT. 

The experiments we performed to demonstrate this framework show that modern models are susceptible to augmentations modeling distribution shifts due to increased CT noise, streaking artifacts caused by high-attenuation implants, and patient movement part way through the scan. More specifically, we show different architectures have differing levels of robustness to these OOD phenomena. We find that nnUnet had higher robustness to increased CT noise and streaking artifacts compared to 3D-Unet, when segmenting the liver. A further experiment, introducing the same suite of training augmentations used in nnUnet to 3D-Unet, shows nnUnet's superior robustness was partially due to the fixed augmentations and partially due to the use of an ensemble of Unets in its training protocol. Furthermore, we find that adding intensity windowing to the pre-processing, as is common for CT image analysis, improves the segmentation performance but reduces the robustness to increased CT noise or metal implant. This is because the windowing removes image detail in the liver and surrounding tissue, while maintaining an intensity boundary between them, making the segmentation task easier for in-distribution images. However, this also means the trained 3D-Unet is less able to handle the intensity variation caused by increased CT noise or streak artifacts, so performance for those OOD test images is worse than if windowing is not used. Experiments assessing the robustness of an object detection architecture, retinaNet, when applied to nodule detection, shows that it is highly robust to increased CT noise. When the 3D-Unet and nnUnet segmentation models are applied, or converted to object detection models, they perform worse than retinaNet without augmentation and also show worse robustness to noise. All models tested show significant weakness to the streaking artifacts caused by metal implants.

We find that the mean model performance decreases with increasing severity level of the noise and metal artifact augmentations, but the variance in the performance for individual cases increases, reflected by positive standard-deviation degradation scores. This has been seen in other research for augmentations used in robustness testing models for MRI segmentation \citep{boone2023rood}. However, we see some effects which are specific to CT artifact augmentations. Abrupt boundaries between adjacent image slices introduced by the augmentations can improve the segmentation or object detection accuracy in some cases while decreasing the average performance across the dataset, thus greatly increasing the variance. This is seen when a motion discontinuity is simulated close to the boundary of the annotated organ or nodule - see Figure \ref{rot_dist_comparison}. This effect is specific to simulating the CT modality because sinograms are acquired slice by slice so the profile of artifacts may spread across an axial plane while abruptly changing from one plane to the next. This contrasts an MRI protocol which reconstructs a 3D image after sampling a 3D trajectory through k-space, resulting in artifacts which propagate in all directions \citep{narai2022movement}.  

In order to empirically scale the weights for our degradation calculations we measured the distribution of CT Noise in the datasets, but a limitation is that we did not do this for the other augmentations. For CT noise, this was achieved by extracting the noise and measuring its standard deviation, which corresponds to how the severity of the noise artifact is quantified. However, for other types of augmentation, this empirical modeling is difficult given the challenges in translating observed image variances to simulation based transforms. For example, streaking artifacts are hard to retrospectively quantify in the image due to its very heterogeneous spacial profile. So instead the severity of the augmentation is defined by the metal implant size, a parameter of the simulation used to generate the artifact. Further studies quantifying the characteristics of distribution shifts of artifacts in CT may enable more specific definition of model robustness, based on degradation weights fitted to the distribution associated with a specific imaging center.

A limitation of the CT simulation emerges for some of the test CT images, which had anatomy cropped by the circular field of view. As the area beyond the field of view is assumed to be air, this would have reduced the apparent attenuation of the X-ray beams at some angles in the simulation compared to a real like CT scan, which may have altered the noise distribution. This could be mitigated in future research by registering a non-cropped atlas CT image to these cases, allowing an approximation of the attenuation beyond the field of view.

A limitation of our motion augmentation method is that it only considers a single rigid movement due to patient changing position, which occurs between the acquisition of individual slices. However, fast motion during the acquisition a sinogram and the impact of complicated artifacts which follow are not considered. In lung imaging, deformable motion due to breathing or the heartbeat during the acquisition of a slice sinogram can cause artifacts which appear as the duplication or distortion of blood vessels and nodules \citep{van2022generating}, which could affect a model more than a rigid discontinuity. In one study radiologists identified artifacts of this type severe enough to affect diagnostic interpretation in 30\% of CT scans acquired from outpatients \citep{dasegowda2023auto}. An augmentation for fast motion artifacts could be added in future work, by applying a deformable motion transform to the image before a subset of the projection angles are sampled during the simulation \citep{van2022generating}. However, determining a corresponding transform for the annotated segmentation maps of object box co-ordinates would be a problem for robustness testing. 

Our presented framework enables direct comparison of black-box models for the same segmentation or object detection task, so a potential user with a limited annotated local dataset can test and compare the robustness of available models against potential CT artifacts or image quality changes which could occur during its use life at their site. The augmentations could also be used to generate a larger benchmarking dataset for the community to study architectural and model-design considerations that result in improved robustness. We recommend that the parameters which determine which slices are affected by simulated CT artifacts, i.e. the timing of motion, the axial position of an implant, or the implant's extension in axial direction, are randomly varied rather than determined by the annotation, to avoid sharp discontinuities between the slices influencing the tested models with implied information about those annotations.




\acks{This research was funded by Innovate UK, grant 10033899. It was further supported by the Wellcome/EPSRC Centre for Medical Engineering [WT 203148/Z/16/Z].}

%
\ethics{This work was based on publicly available datasets, see below.}

\coi{Julia A. Schnabel: Founder of XRnostics Ltd. \\ Kanwal K. Bhatia: Founder of Metalynx Ltd. trading as Aival. \\ Jack Highton, Quok Zong Chong, Samuel Finestone, and Arian Beqiri: employees of Metalynx Ltd. trading as Aival.}
\newpage
\data{This work was based on publicly available datasets. Luna 16: https://luna16.grand-challenge.org/ Segmentation Decathlon: http://medicaldecathlon.com/}

\bibliography{sample}

\begin{thebibliography}{72}
\providecommand{\natexlab}[1]{#1}
\providecommand{\url}[1]{\texttt{#1}}
\expandafter\ifx\csname urlstyle\endcsname\relax
  \providecommand{\doi}[1]{doi: #1}\else
  \providecommand{\doi}{doi: \begingroup \urlstyle{rm}\Url}\fi

\bibitem[Ahmad(2021)]{ahmad2021reviewing}
Rani Ahmad.
\newblock Reviewing the relationship between machines and radiology: the application of artificial intelligence.
\newblock \emph{Acta Radiologica Open}, 10\penalty0 (2):\penalty0 2058460121990296, 2021.

\bibitem[Anthony and Kamnitsas(2023)]{anthony2023use}
Harry Anthony and Konstantinos Kamnitsas.
\newblock On the use of mahalanobis distance for out-of-distribution detection with neural networks for medical imaging.
\newblock In \emph{International Workshop on Uncertainty for Safe Utilization of Machine Learning in Medical Imaging}, pages 136--146. Springer, 2023.

\bibitem[Antonelli et~al.(2022)Antonelli, Reinke, Bakas, Farahani, Kopp-Schneider, Landman, Litjens, Menze, Ronneberger, Summers, et~al.]{antonelli2022medical}
Michela Antonelli, Annika Reinke, Spyridon Bakas, Keyvan Farahani, Annette Kopp-Schneider, Bennett~A Landman, Geert Litjens, Bjoern Menze, Olaf Ronneberger, Ronald~M Summers, et~al.
\newblock The medical segmentation decathlon.
\newblock \emph{Nature Communications}, 13\penalty0 (1):\penalty0 4128, 2022.

\bibitem[Armato~III et~al.(2011)Armato~III, McLennan, Bidaut, McNitt-Gray, Meyer, Reeves, Zhao, Aberle, Henschke, Hoffman, et~al.]{armato2011lung}
Samuel~G Armato~III, Geoffrey McLennan, Luc Bidaut, Michael~F McNitt-Gray, Charles~R Meyer, Anthony~P Reeves, Binsheng Zhao, Denise~R Aberle, Claudia~I Henschke, Eric~A Hoffman, et~al.
\newblock The lung image database consortium (lidc) and image database resource initiative (idri): a completed reference database of lung nodules on ct scans.
\newblock \emph{Medical Physics}, 38\penalty0 (2):\penalty0 915--931, 2011.

\bibitem[Barrett and Keat(2004)]{barrett2004artifacts}
Julia~F Barrett and Nicholas Keat.
\newblock Artifacts in ct: recognition and avoidance.
\newblock \emph{Radiographics}, 24\penalty0 (6):\penalty0 1679--1691, 2004.

\bibitem[Bilic et~al.(2023)Bilic, Christ, Li, Vorontsov, Ben-Cohen, Kaissis, Szeskin, Jacobs, Mamani, Chartrand, et~al.]{bilic2023liver}
Patrick Bilic, Patrick Christ, Hongwei~Bran Li, Eugene Vorontsov, Avi Ben-Cohen, Georgios Kaissis, Adi Szeskin, Colin Jacobs, Gabriel Efrain~Humpire Mamani, Gabriel Chartrand, et~al.
\newblock The liver tumor segmentation benchmark (lits).
\newblock \emph{Medical Image Analysis}, 84:\penalty0 102680, 2023.

\bibitem[Boedeker et~al.(2007)Boedeker, Cooper, and McNitt-Gray]{boedeker2007application}
Kirsten~L Boedeker, Virgil~N Cooper, and Michael~F McNitt-Gray.
\newblock Application of the noise power spectrum in modern diagnostic mdct: part i. measurement of noise power spectra and noise equivalent quanta.
\newblock \emph{Physics in Medicine \& biology}, 52\penalty0 (14):\penalty0 4027, 2007.

\bibitem[Bolliger et~al.(2009)Bolliger, Oesterhelweg, Spendlove, Ross, and Thali]{bolliger2009differentiation}
Stephan~A Bolliger, Lars Oesterhelweg, Danny Spendlove, Steffen Ross, and Michael~J Thali.
\newblock Is differentiation of frequently encountered foreign bodies in corpses possible by hounsfield density measurement?
\newblock \emph{Journal of Forensic Sciences}, 54\penalty0 (5):\penalty0 1119--1122, 2009.

\bibitem[Boone et~al.(2023)Boone, Biparva, Forooshani, Ramirez, Masellis, Bartha, Symons, Strother, Black, Heyn, et~al.]{boone2023rood}
Lyndon Boone, Mahdi Biparva, Parisa~Mojiri Forooshani, Joel Ramirez, Mario Masellis, Robert Bartha, Sean Symons, Stephen Strother, Sandra~E Black, Chris Heyn, et~al.
\newblock Rood-mri: Benchmarking the robustness of deep learning segmentation models to out-of-distribution and corrupted data in mri.
\newblock \emph{NeuroImage}, 278:\penalty0 120289, 2023.

\bibitem[Bui et~al.(2019)Bui, Wang, Chen, Lin, Li, and Shen]{bui2019multi}
Toan~Duc Bui, Li~Wang, Jian Chen, Weili Lin, Gang Li, and Dinggang Shen.
\newblock Multi-task learning for neonatal brain segmentation using 3d dense-unet with dense attention guided by geodesic distance.
\newblock In \emph{Domain Adaptation and Representation Transfer and Medical Image Learning with Less Labels and Imperfect Data: First MICCAI Workshop, DART 2019, and First International Workshop, MIL3ID 2019, Shenzhen, Held in Conjunction with MICCAI 2019, Shenzhen, China, October 13 and 17, 2019, Proceedings 1}, pages 243--251. Springer, 2019.

\bibitem[Cabral~Jr. et~al.(1993)Cabral~Jr., White, Kim, and Effmann]{cabral1993interactive}
James~E Cabral~Jr., Keith~S White, Yongmin Kim, and Eric~L Effmann.
\newblock Interactive segmentation of brain tumors in mr images using 3d region growing.
\newblock In \emph{Medical Imaging 1993: Image Processing}, volume 1898, pages 171--181. SPIE, 1993.

\bibitem[Campello et~al.(2021)Campello, Gkontra, Izquierdo, Martin-Isla, Sojoudi, Full, Maier-Hein, Zhang, He, Ma, et~al.]{campello2021multi}
Victor~M Campello, Polyxeni Gkontra, Cristian Izquierdo, Carlos Martin-Isla, Alireza Sojoudi, Peter~M Full, Klaus Maier-Hein, Yao Zhang, Zhiqiang He, Jun Ma, et~al.
\newblock Multi-centre, multi-vendor and multi-disease cardiac segmentation: the m\&ms challenge.
\newblock \emph{IEEE Transactions on Medical Imaging}, 40\penalty0 (12):\penalty0 3543--3554, 2021.

\bibitem[Cardoso et~al.(2022)Cardoso, Li, Brown, Ma, Kerfoot, Wang, Murrey, Myronenko, Zhao, Yang, et~al.]{cardoso2022monai}
M~Jorge Cardoso, Wenqi Li, Richard Brown, Nic Ma, Eric Kerfoot, Yiheng Wang, Benjamin Murrey, Andriy Myronenko, Can Zhao, Dong Yang, et~al.
\newblock {MONAI}: An open-source framework for deep learning in healthcare.
\newblock \emph{arXiv preprint arXiv:2211.02701}, 2022.

\bibitem[Choi et~al.(2022)Choi, Cho, Ha, Lee, Koh, Seo, Choi, Cheon, Phi, Kim, et~al.]{choi2022deep}
Jae~Won Choi, Yeon~Jin Cho, Ji~Young Ha, Yun~Young Lee, Seok~Young Koh, June~Young Seo, Young~Hun Choi, Jung-Eun Cheon, Ji~Hoon Phi, Injoon Kim, et~al.
\newblock Deep learning-assisted diagnosis of pediatric skull fractures on plain radiographs.
\newblock \emph{Korean Journal of Radiology}, 23\penalty0 (3):\penalty0 343, 2022.

\bibitem[{\c{C}}i{\c{c}}ek et~al.(2016){\c{C}}i{\c{c}}ek, Abdulkadir, Lienkamp, Brox, and Ronneberger]{cciccek20163d}
{\"O}zg{\"u}n {\c{C}}i{\c{c}}ek, Ahmed Abdulkadir, Soeren~S Lienkamp, Thomas Brox, and Olaf Ronneberger.
\newblock {3D U-N}et: learning dense volumetric segmentation from sparse annotation.
\newblock In \emph{Medical Image Computing and Computer-Assisted Intervention--MICCAI 2016: 19th International Conference, Athens, Greece, October 17-21, 2016, Proceedings, Part II 19}, pages 424--432. Springer, 2016.

\bibitem[Dasegowda et~al.(2023)Dasegowda, Bizzo, Kaviani, Karout, Ebrahimian, Digumarthy, Neumark, Hillis, Kalra, and Dreyer]{dasegowda2023auto}
Giridhar Dasegowda, Bernardo~C Bizzo, Parisa Kaviani, Lina Karout, Shadi Ebrahimian, Subba~R Digumarthy, Nir Neumark, James~M Hillis, Mannudeep~K Kalra, and Keith~J Dreyer.
\newblock Auto-detection of motion artifacts on ct pulmonary angiograms with a physician-trained ai algorithm.
\newblock \emph{Diagnostics}, 13\penalty0 (4):\penalty0 778, 2023.

\bibitem[Dice(1945)]{dice1945measures}
Lee~R Dice.
\newblock Measures of the amount of ecologic association between species.
\newblock \emph{Ecology}, 26\penalty0 (3):\penalty0 297--302, 1945.

\bibitem[Dolly et~al.(2016)Dolly, Chen, Anastasio, Mutic, and Li]{dolly2016practical}
Steven Dolly, Hsin-Chen Chen, Mark Anastasio, Sasa Mutic, and Hua Li.
\newblock Practical considerations for noise power spectra estimation for clinical ct scanners.
\newblock \emph{Journal of Applied Clinical Medical Physics}, 17\penalty0 (3):\penalty0 392--407, 2016.

\bibitem[Galati et~al.(2022)Galati, Ourselin, and Zuluaga]{galati2022accuracy}
Francesco Galati, S{\'e}bastien Ourselin, and Maria~A Zuluaga.
\newblock From accuracy to reliability and robustness in cardiac magnetic resonance image segmentation: a review.
\newblock \emph{Applied Sciences}, 12\penalty0 (8):\penalty0 3936, 2022.

\bibitem[Geirhos et~al.(2018)Geirhos, Temme, Rauber, Sch{\"u}tt, Bethge, and Wichmann]{geirhos2018generalisation}
Robert Geirhos, Carlos~RM Temme, Jonas Rauber, Heiko~H Sch{\"u}tt, Matthias Bethge, and Felix~A Wichmann.
\newblock Generalisation in humans and deep neural networks.
\newblock \emph{Advances in Neural Information Processing Systems}, 31, 2018.

\bibitem[Giorgio and De~Stefano(2013)]{giorgio2013clinical}
Antonio Giorgio and Nicola De~Stefano.
\newblock Clinical use of brain volumetry.
\newblock \emph{Journal of Magnetic Resonance Imaging}, 37\penalty0 (1):\penalty0 1--14, 2013.

\bibitem[Goceri(2023)]{goceri2023medical}
Evgin Goceri.
\newblock Medical image data augmentation: techniques, comparisons and interpretations.
\newblock \emph{Artificial Intelligence Review}, pages 1--45, 2023.

\bibitem[Goldman(2007)]{goldman2007principles}
Lee~W Goldman.
\newblock Principles of ct: dose and image quality ct.
\newblock \emph{Journal of Nuclear Medicine Technology}, 35\penalty0 (4):\penalty0 213--225, 2007.

\bibitem[Hendrycks and Gimpel(2016)]{hendrycks2016baseline}
Dan Hendrycks and Kevin Gimpel.
\newblock A baseline for detecting misclassified and out-of-distribution examples in neural networks.
\newblock In \emph{International Conference on Learning Representations}, 2016.

\bibitem[Hermena and Young(2021)]{hermena2021ct}
Shady Hermena and Michael Young.
\newblock \emph{CT-scan image production procedures}.
\newblock StatPearls. Treasure Island, (FL): StatPearls, 2021.
\newblock PMID: 34662062.

\bibitem[Isensee et~al.(2021)Isensee, Jaeger, Kohl, Petersen, and Maier-Hein]{isensee2021nnu}
Fabian Isensee, Paul~F Jaeger, Simon~AA Kohl, Jens Petersen, and Klaus~H Maier-Hein.
\newblock nn{U-N}et: a self-configuring method for deep learning-based biomedical image segmentation.
\newblock \emph{Nature methods}, 18\penalty0 (2):\penalty0 203--211, 2021.

\bibitem[Jacobs and van Ginneken(2016)]{luna16web}
Colin Jacobs and Bram van Ginneken.
\newblock Lung nodule analysis 2016, 2016.
\newblock URL \url{https://luna16.grand-challenge.org/}.

\bibitem[Jacobson and Krupinski(2021)]{jacobson2021clinical}
Francine~L Jacobson and Elizabeth~A Krupinski.
\newblock Clinical validation is the key to adopting ai in clinical practice.
\newblock \emph{Radiology: Artificial Intelligence}, 3\penalty0 (4):\penalty0 e210104, 2021.

\bibitem[Jin et~al.(2022)Jin, Udupa, Zhao, Tong, Odhner, Pednekar, Nag, Lewis, Poole, Mannikeri, et~al.]{jin2022object}
Chao Jin, Jayaram~K Udupa, Liming Zhao, Yubing Tong, Dewey Odhner, Gargi Pednekar, Sanghita Nag, Sharon Lewis, Nicholas Poole, Sutirth Mannikeri, et~al.
\newblock Object recognition in medical images via anatomy-guided deep learning.
\newblock \emph{Medical Image Analysis}, 81:\penalty0 102527, 2022.

\bibitem[Khalifa et~al.(2022)Khalifa, Loey, and Mirjalili]{khalifa2022comprehensive}
Nour~Eldeen Khalifa, Mohamed Loey, and Seyedali Mirjalili.
\newblock A comprehensive survey of recent trends in deep learning for digital images augmentation.
\newblock \emph{Artificial Intelligence Review}, pages 1--27, 2022.

\bibitem[Koh et~al.(2022)Koh, Papanikolaou, Bick, Illing, Kahn~Jr., Kalpathi-Cramer, Matos, Mart{\'\i}-Bonmat{\'\i}, Miles, Mun, et~al.]{koh2022artificial}
Dow-Mu Koh, Nickolas Papanikolaou, Ulrich Bick, Rowland Illing, Charles~E Kahn~Jr., Jayshree Kalpathi-Cramer, Celso Matos, Luis Mart{\'\i}-Bonmat{\'\i}, Anne Miles, Seong~Ki Mun, et~al.
\newblock Artificial intelligence and machine learning in cancer imaging.
\newblock \emph{Communications Medicine}, 2\penalty0 (1):\penalty0 133, 2022.

\bibitem[Kooi et~al.(2017)Kooi, Litjens, Van~Ginneken, Gubern-M{\'e}rida, S{\'a}nchez, Mann, den Heeten, and Karssemeijer]{kooi2017large}
Thijs Kooi, Geert Litjens, Bram Van~Ginneken, Albert Gubern-M{\'e}rida, Clara~I S{\'a}nchez, Ritse Mann, Ard den Heeten, and Nico Karssemeijer.
\newblock Large scale deep learning for computer aided detection of mammographic lesions.
\newblock \emph{Medical Image Analysis}, 35:\penalty0 303--312, 2017.

\bibitem[Larici et~al.(2017)Larici, Farchione, Franchi, Ciliberto, Cicchetti, Calandriello, Del~Ciello, and Bonomo]{larici2017lung}
Anna~Rita Larici, Alessandra Farchione, Paola Franchi, Mario Ciliberto, Giuseppe Cicchetti, Lucio Calandriello, Annemilia Del~Ciello, and Lorenzo Bonomo.
\newblock Lung nodules: size still matters.
\newblock \emph{European Respiratory Review}, 26\penalty0 (146), 2017.

\bibitem[Li et~al.(2023{\natexlab{a}})Li, Liu, Fan, Li, and Zhang]{li2023self}
Wei Li, Guanghai Liu, Haoyi Fan, Zuoyong Li, and David Zhang.
\newblock Self-supervised multi-scale cropping and simple masked attentive predicting for lung ct-scan anomaly detection.
\newblock \emph{IEEE Transactions on Medical Imaging}, 2023{\natexlab{a}}.

\bibitem[Li et~al.(2023{\natexlab{b}})Li, Lin, Zhang, Feng, Huang, and Bai]{li2023automatic}
Yuchun Li, Cong Lin, Yu~Zhang, Siling Feng, Mengxing Huang, and Zhiming Bai.
\newblock Automatic segmentation of prostate mri based on 3d pyramid pooling unet.
\newblock \emph{Medical Physics}, 50\penalty0 (2):\penalty0 906--921, 2023{\natexlab{b}}.

\bibitem[Liguori et~al.(2015)Liguori, Frauenfelder, Massaroni, Saccomandi, Giurazza, Pitocco, Marano, and Schena]{liguori2015emerging}
Carlo Liguori, Giulia Frauenfelder, Carlo Massaroni, Paola Saccomandi, Francesco Giurazza, Francesca Pitocco, Riccardo Marano, and Emiliano Schena.
\newblock Emerging clinical applications of computed tomography.
\newblock \emph{Medical Devices: Evidence and Research}, pages 265--278, 2015.

\bibitem[Lin et~al.(2017)Lin, Goyal, Girshick, He, and Doll{\'a}r]{lin2017focal}
Tsung-Yi Lin, Priya Goyal, Ross Girshick, Kaiming He, and Piotr Doll{\'a}r.
\newblock Focal loss for dense object detection.
\newblock In \emph{Proceedings of the IEEE International Conference on Computer Vision}, pages 2980--2988, 2017.

\bibitem[Liu et~al.(2020)Liu, Wang, Owens, and Li]{liu2020energy}
Weitang Liu, Xiaoyun Wang, John Owens, and Yixuan Li.
\newblock Energy-based out-of-distribution detection.
\newblock \emph{Advances in Neural Information Processing Systems}, 33:\penalty0 21464--21475, 2020.

\bibitem[Liu et~al.(2022)Liu, Liang, Deng, Tan, and Xie]{liu2022learning}
Xuan Liu, Xiaokun Liang, Lei Deng, Shan Tan, and Yaoqin Xie.
\newblock Learning low-dose ct degradation from unpaired data with flow-based model.
\newblock \emph{Medical Physics}, 49\penalty0 (12):\penalty0 7516--7530, 2022.

\bibitem[Maier-Hein et~al.(2024)Maier-Hein, Reinke, Godau, Tizabi, Buettner, Christodoulou, Glocker, Isensee, Kleesiek, Kozubek, et~al.]{maier2024metrics}
Lena Maier-Hein, Annika Reinke, Patrick Godau, Minu~D Tizabi, Florian Buettner, Evangelia Christodoulou, Ben Glocker, Fabian Isensee, Jens Kleesiek, Michal Kozubek, et~al.
\newblock Metrics reloaded: recommendations for image analysis validation.
\newblock \emph{Nature Methods}, pages 1--18, 2024.

\bibitem[Medawar et~al.(2021)Medawar, Thieleking, Manuilova, Paerisch, Villringer, Witte, and Beyer]{medawar2021estimating}
Evelyn Medawar, Ronja Thieleking, Iryna Manuilova, Maria Paerisch, Arno Villringer, A~Veronica Witte, and Frauke Beyer.
\newblock Estimating the effect of a scanner upgrade on measures of grey matter structure for longitudinal designs.
\newblock \emph{PLOS One}, 16\penalty0 (10):\penalty0 e0239021, 2021.

\bibitem[N{\'a}rai et~al.(2022)N{\'a}rai, Hermann, Auer, Kemenczky, Szalma, Homolya, Somogyi, Vakli, Weiss, and Vidny{\'a}nszky]{narai2022movement}
{\'A}d{\'a}m N{\'a}rai, Petra Hermann, Tibor Auer, P{\'e}ter Kemenczky, J{\'a}nos Szalma, Istv{\'a}n Homolya, Eszter Somogyi, P{\'a}l Vakli, B{\'e}la Weiss, and Zolt{\'a}n Vidny{\'a}nszky.
\newblock Movement-related artefacts (mr-art) dataset of matched motion-corrupted and clean structural mri brain scans.
\newblock \emph{Scientific Data}, 9\penalty0 (1):\penalty0 630, 2022.

\bibitem[Nguyen et~al.(2023)Nguyen, Pham, Diep, Phan, Pham, Tong, Nguyen, Le, Ho, Xie, et~al.]{nguyen2023out}
Duy Minh~Ho Nguyen, Tan~Ngoc Pham, Nghiem~Tuong Diep, Nghi~Quoc Phan, Quang Pham, Vinh Tong, Binh~T Nguyen, Ngan~Hoang Le, Nhat Ho, Pengtao Xie, et~al.
\newblock On the out of distribution robustness of foundation models in medical image segmentation.
\newblock In \emph{Advances in Neural Information Processing Systems 36 (NeurIPS 2023): R0-FoMo: Robustness of Few-shot and Zero-shot Learning in Large Foundation Models}, 2023.

\bibitem[Noguchi et~al.(2020)Noguchi, Nishio, Yakami, Nakagomi, and Togashi]{noguchi2020bone}
Shunjiro Noguchi, Mizuho Nishio, Masahiro Yakami, Keita Nakagomi, and Kaori Togashi.
\newblock Bone segmentation on whole-body ct using convolutional neural network with novel data augmentation techniques.
\newblock \emph{Computers in Biology and Medicine}, 121:\penalty0 103767, 2020.

\bibitem[Omigbodun et~al.(2019)Omigbodun, Noo, McNitt-Gray, Hsu, and Hsieh]{omigbodun2019effects}
Akinyinka~O Omigbodun, Frederic Noo, Michael McNitt-Gray, William Hsu, and Scott~S Hsieh.
\newblock The effects of physics-based data augmentation on the generalizability of deep neural networks: Demonstration on nodule false-positive reduction.
\newblock \emph{Medical Physics}, 46\penalty0 (10):\penalty0 4563--4574, 2019.

\bibitem[Padilla et~al.(2020)Padilla, Netto, and Da~Silva]{padilla2020survey}
Rafael Padilla, Sergio~L Netto, and Eduardo~AB Da~Silva.
\newblock A survey on performance metrics for object-detection algorithms.
\newblock In \emph{2020 International Conference on Systems, Signals and Image Processing (IWSSIP)}, pages 237--242. IEEE, 2020.

\bibitem[Padilla et~al.(2021)Padilla, Passos, Dias, Netto, and Da~Silva]{padilla2021comparative}
Rafael Padilla, Wesley~L Passos, Thadeu~LB Dias, Sergio~L Netto, and Eduardo~AB Da~Silva.
\newblock A comparative analysis of object detection metrics with a companion open-source toolkit.
\newblock \emph{Electronics}, 10\penalty0 (3):\penalty0 279, 2021.

\bibitem[Peyrin and Engelke(2021)]{peyrin2021ct}
Fran{\c{c}}oise Peyrin and Klaus Engelke.
\newblock Ct imaging: Basics and new trends.
\newblock In \emph{Handbook of Particle Detection and Imaging}, pages 1173--1215. Springer, 2021.

\bibitem[Potvin et~al.(2019)Potvin, Khademi, Chouinard, Farokhian, Dieumegarde, Leppert, Hoge, Rajah, Bellec, Duchesne, et~al.]{potvin2019measurement}
Olivier Potvin, April Khademi, Isabelle Chouinard, Farnaz Farokhian, Louis Dieumegarde, Ilana Leppert, Rick Hoge, Maria~Natasha Rajah, Pierre Bellec, Simon Duchesne, et~al.
\newblock Measurement variability following mri system upgrade.
\newblock \emph{Frontiers in Neurology}, 10:\penalty0 726, 2019.

\bibitem[Prior et~al.(2020)Prior, Almeida, Kathiravelu, Kurc, Smith, Fitzgerald, and Saltz]{prior2020open}
Fred Prior, J~Almeida, P~Kathiravelu, T~Kurc, K~Smith, Thomas~J Fitzgerald, and J~Saltz.
\newblock Open access image repositories: high-quality data to enable machine learning research.
\newblock \emph{Clinical Radiology}, 75\penalty0 (1):\penalty0 7--12, 2020.

\bibitem[Qui{\~n}onero-Candela et~al.(2008)Qui{\~n}onero-Candela, Sugiyama, Schwaighofer, and Lawrence]{quinonero2008dataset}
Joaquin Qui{\~n}onero-Candela, Masashi Sugiyama, Anton Schwaighofer, and Neil~D Lawrence.
\newblock \emph{Dataset shift in machine learning}.
\newblock MIT Press, 2008.

\bibitem[Ramachandran and Lakshminarayanan(1971)]{ramachandran1971three}
GN~Ramachandran and AV~Lakshminarayanan.
\newblock Three-dimensional reconstruction from radiographs and electron micrographs: application of convolutions instead of fourier transforms.
\newblock \emph{Proceedings of the National Academy of Sciences}, 68\penalty0 (9):\penalty0 2236--2240, 1971.

\bibitem[Ren et~al.(2020)Ren, Zhou, Liu, Peng, Shi, Xu, Shan, and Liu]{ren2020unsupervised}
He~Ren, Lingxiao Zhou, Gang Liu, Xueqing Peng, Weiya Shi, Huilin Xu, Fei Shan, and Lei Liu.
\newblock An unsupervised semi-automated pulmonary nodule segmentation method based on enhanced region growing.
\newblock \emph{Quantitative Imaging in Medicine and Surgery}, 10\penalty0 (1):\penalty0 233, 2020.

\bibitem[Rubin(1992)]{rubin1992nonlinenr}
LI~Rubin.
\newblock Nonlinenr total variation based noise removal algorithms.
\newblock \emph{Physica D: Nonlinear Phenomena}, 60:\penalty0 259--265, 1992.

\bibitem[Sanaat et~al.(2022)Sanaat, Shiri, Ferdowsi, Arabi, and Zaidi]{sanaat2022robust}
Amirhossein Sanaat, Isaac Shiri, Sohrab Ferdowsi, Hossein Arabi, and Habib Zaidi.
\newblock Robust-deep: a method for increasing brain imaging datasets to improve deep learning models’ performance and robustness.
\newblock \emph{Journal of Digital Imaging}, 35\penalty0 (3):\penalty0 469--481, 2022.

\bibitem[Shaikhina and Khovanova(2017)]{shaikhina2017handling}
Torgyn Shaikhina and Natalia~A Khovanova.
\newblock Handling limited datasets with neural networks in medical applications: A small-data approach.
\newblock \emph{Artificial Intelligence in Medicine}, 75:\penalty0 51--63, 2017.

\bibitem[Shaw et~al.(2020)Shaw, Sudre, Varsavsky, Ourselin, and Cardoso]{shaw2020k}
Richard Shaw, Carole~H Sudre, Thomas Varsavsky, S{\'e}bastien Ourselin, and M~Jorge Cardoso.
\newblock A k-space model of movement artefacts: application to segmentation augmentation and artefact removal.
\newblock \emph{IEEE Transactions on Medical Imaging}, 39\penalty0 (9):\penalty0 2881--2892, 2020.

\bibitem[Singla et~al.(2022)Singla, Ringstrom, Hu, Lessoway, Reid, Rohling, and Nguan]{singla2022speckle}
Rohit Singla, Cailin Ringstrom, Ricky Hu, Victoria Lessoway, Janice Reid, Robert Rohling, and Christophe Nguan.
\newblock Speckle and shadows: ultrasound-specific physics-based data augmentation for kidney segmentation.
\newblock In \emph{International Conference on Medical Imaging with Deep Learning}, pages 1139--1148. PMLR, 2022.

\bibitem[Szegedy et~al.(2013)Szegedy, Zaremba, Sutskever, Bruna, Erhan, Goodfellow, and Fergus]{szegedy2013intriguing}
Christian Szegedy, Wojciech Zaremba, Ilya Sutskever, Joan Bruna, Dumitru Erhan, Ian Goodfellow, and Rob Fergus.
\newblock Intriguing properties of neural networks.
\newblock \emph{arXiv preprint arXiv:1312.6199}, 2013.

\bibitem[Van~Aarle et~al.(2016)Van~Aarle, Palenstijn, Cant, Janssens, Bleichrodt, Dabravolski, De~Beenhouwer, Batenburg, and Sijbers]{van2016fast}
Wim Van~Aarle, Willem~Jan Palenstijn, Jeroen Cant, Eline Janssens, Folkert Bleichrodt, Andrei Dabravolski, Jan De~Beenhouwer, K~Joost Batenburg, and Jan Sijbers.
\newblock Fast and flexible x-ray tomography using the astra toolbox.
\newblock \emph{Optics Express}, 24\penalty0 (22):\penalty0 25129--25147, 2016.

\bibitem[van~der Ham et~al.(2022)van~der Ham, Latisenko, Tsiaousis, and van Tulder]{van2022generating}
Guus van~der Ham, Rudolfs Latisenko, Michail Tsiaousis, and Gijs van Tulder.
\newblock Generating artificial artifacts for motion artifact detection in chest ct.
\newblock In \emph{International Workshop on Simulation and Synthesis in Medical Imaging}, pages 12--23. Springer, 2022.

\bibitem[Vasiliuk et~al.(2023{\natexlab{a}})Vasiliuk, Frolova, Belyaev, and Shirokikh]{vasiliuk2023limitations}
Anton Vasiliuk, Daria Frolova, Mikhail Belyaev, and Boris Shirokikh.
\newblock Limitations of out-of-distribution detection in 3d medical image segmentation.
\newblock \emph{Journal of Imaging}, 9\penalty0 (9):\penalty0 191, 2023{\natexlab{a}}.

\bibitem[Vasiliuk et~al.(2023{\natexlab{b}})Vasiliuk, Frolova, Belyaev, and Shirokikh]{vasiliuk2023redesigning}
Anton Vasiliuk, Daria Frolova, Mikhail Belyaev, and Boris Shirokikh.
\newblock Redesigning out-of-distribution detection on 3d medical images.
\newblock In \emph{International Workshop on Uncertainty for Safe Utilization of Machine Learning in Medical Imaging}, pages 126--135. Springer, 2023{\natexlab{b}}.

\bibitem[Vayena et~al.(2018)Vayena, Blasimme, and Cohen]{vayena2018machine}
Effy Vayena, Alessandro Blasimme, and I~Glenn Cohen.
\newblock Machine learning in medicine: addressing ethical challenges.
\newblock \emph{PLOS Medicine}, 15\penalty0 (11):\penalty0 e1002689, 2018.

\bibitem[Waheed et~al.(2020)Waheed, Goyal, Gupta, Khanna, Al-Turjman, and Pinheiro]{waheed2020covidgan}
Abdul Waheed, Muskan Goyal, Deepak Gupta, Ashish Khanna, Fadi Al-Turjman, and Pl{\'a}cido~Rogerio Pinheiro.
\newblock Covidgan: data augmentation using auxiliary classifier gan for improved covid-19 detection.
\newblock \emph{Ieee Access}, 8:\penalty0 91916--91923, 2020.

\bibitem[Won~Kim and Kim(2014)]{won2014realistic}
Chang Won~Kim and Jong~Hyo Kim.
\newblock Realistic simulation of reduced-dose ct with noise modeling and sinogram synthesis using dicom ct images.
\newblock \emph{Medical Physics}, 41\penalty0 (1):\penalty0 011901, 2014.

\bibitem[Yang(2023)]{monai_nnunetv2}
Dong Yang.
\newblock {MONAI} and nn{U-N}et integration, 2023.
\newblock URL \url{https://github.com/Project-MONAI/tutorials/tree/main/nnunet}.

\bibitem[Yang et~al.(2023)Yang, Cai, Liu, Yuan, Ma, Chen, Zhang, Wu, and Ge]{yang2023imaging}
Wei Yang, Zecheng Cai, Xiaoyin Liu, Wenqi Yuan, Rong Ma, Zhen Chen, Jianqun Zhang, Peng Wu, and Zhaohui Ge.
\newblock Imaging study of the effect of postural changes on the retroperitoneal oblique corridor in degenerative lumbar scoliosis.
\newblock \emph{European Spine Journal}, pages 1--7, 2023.

\bibitem[Yu et~al.(2022)Yu, Anakwenze, Zhao, Martin, Ludmir, S.~Niedzielski, Qureshi, Das, Holliday, Raldow, et~al.]{yu2022multi}
Cenji Yu, Chidinma~P Anakwenze, Yao Zhao, Rachael~M Martin, Ethan~B Ludmir, Joshua S.~Niedzielski, Asad Qureshi, Prajnan Das, Emma~B Holliday, Ann~C Raldow, et~al.
\newblock Multi-organ segmentation of abdominal structures from non-contrast and contrast enhanced ct images.
\newblock \emph{Scientific Reports}, 12\penalty0 (1):\penalty0 19093, 2022.

\bibitem[Zhou et~al.(2023)Zhou, Gao, Wu, Grundy, Chen, Chen, and Li]{zhou2023modelobfuscator}
Mingyi Zhou, Xiang Gao, Jing Wu, John Grundy, Xiao Chen, Chunyang Chen, and Li~Li.
\newblock Modelobfuscator: Obfuscating model information to protect deployed ml-based systems.
\newblock In \emph{Proceedings of the 32nd ACM SIGSOFT International Symposium on Software Testing and Analysis}, pages 1005--1017, 2023.

\bibitem[Zijdenbos et~al.(1994)Zijdenbos, Dawant, Margolin, and Palmer]{zijdenbos1994morphometric}
Alex~P Zijdenbos, Benoit~M Dawant, Richard~A Margolin, and Andrew~C Palmer.
\newblock Morphometric analysis of white matter lesions in mr images: method and validation.
\newblock \emph{IEEE transactions on medical imaging}, 13\penalty0 (4):\penalty0 716--724, 1994.

\bibitem[Zimmerer et~al.(2022)Zimmerer, Full, Isensee, J{\"a}ger, Adler, Petersen, K{\"o}hler, Ross, Reinke, Kascenas, et~al.]{zimmerer2022mood}
David Zimmerer, Peter~M Full, Fabian Isensee, Paul J{\"a}ger, Tim Adler, Jens Petersen, Gregor K{\"o}hler, Tobias Ross, Annika Reinke, Antanas Kascenas, et~al.
\newblock Mood 2020: A public benchmark for out-of-distribution detection and localization on medical images.
\newblock \emph{IEEE Transactions on Medical Imaging}, 41\penalty0 (10):\penalty0 2728--2738, 2022.

\end{thebibliography}


\clearpage
\appendix

\section{}

        \begin{figure}[H]
		\centering
		\includegraphics[width=1.0\linewidth]{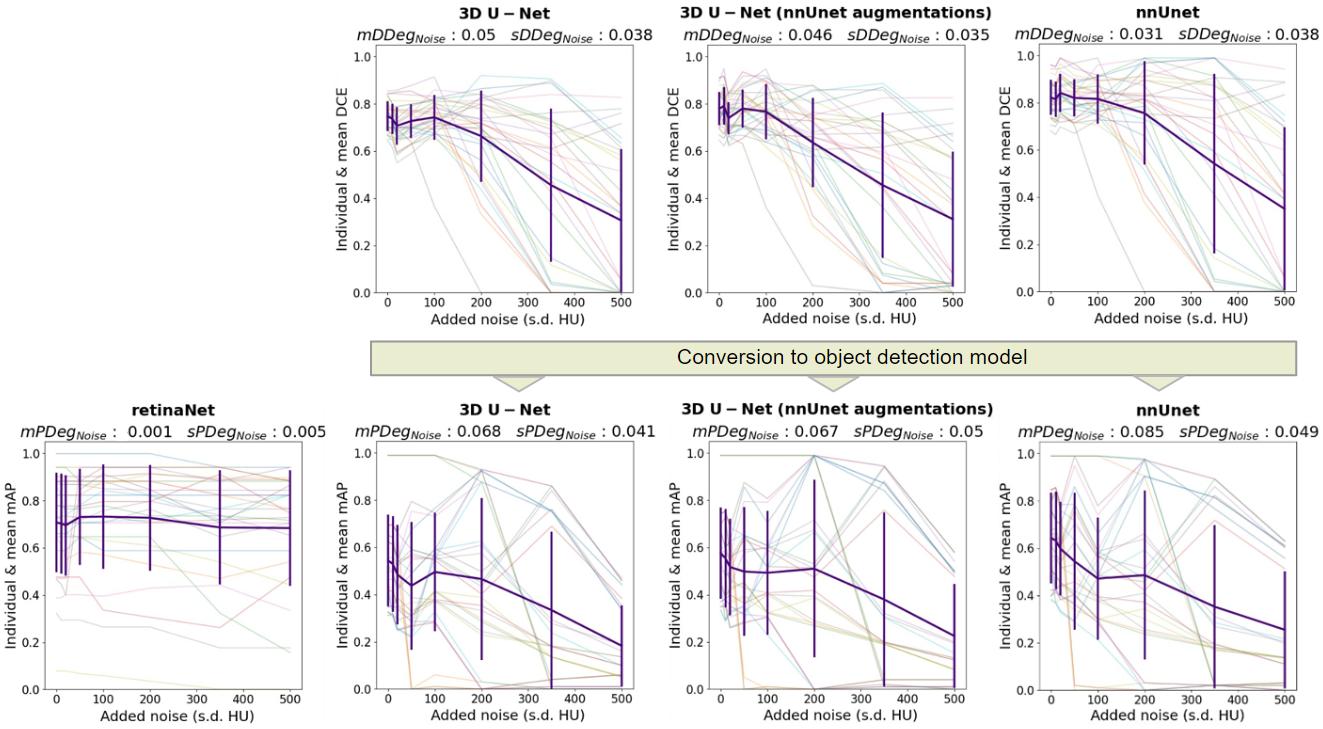}
	\caption{Top: The Dice (DSC) score results against the level of CT noise added by augmentation, for the tested segmentation models. Bottom: mean Average precision (mAP) results against the level of CT noise added by augmentation, for the tested object detection models including segmentation models converted to generate bounding boxes. These are shown for each of the 30 test cases from the LUNA16 dataset with different levels of CT noise added by augmentation. The corresponding means across the test cases is shown in purple, with the error bars representing the standard deviation. The results in this plot are summarized in Tables \ref{tab_dce} and \ref{tab_map}.}
    \label{luna noise seg 7}
    \end{figure}

        \begin{figure}[H]
		\centering
		\includegraphics[width=1.0\linewidth]{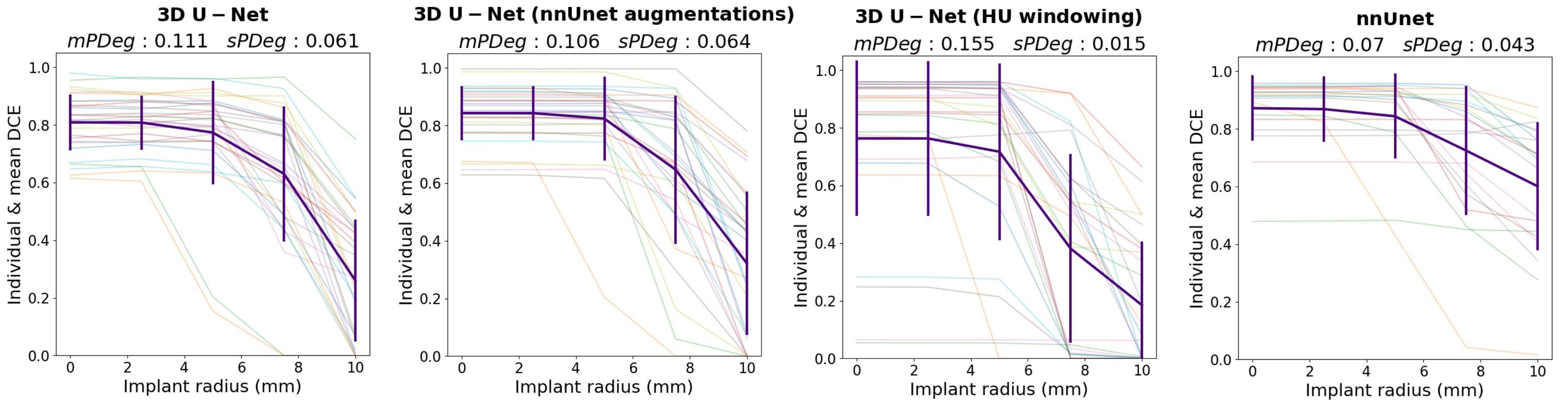}
	\caption{The Dice (DSC) score results for the tested segmentation models, shown for each of the 30 test cases from the liver segmentation dataset, with simulated CT streak artifacts of varying severity caused by augmentation by insertion of a cylindrical spinal implant with varying radius. The corresponding means across the test cases is shown in purple, with the error bars representing the standard deviation. The results in this plot are summarized in Tables \ref{tab_dce}.}
    \label{liver cycl implant 3}
    \end{figure}

    \begin{figure}[H]
		\centering
		\includegraphics[width=1.0\linewidth]{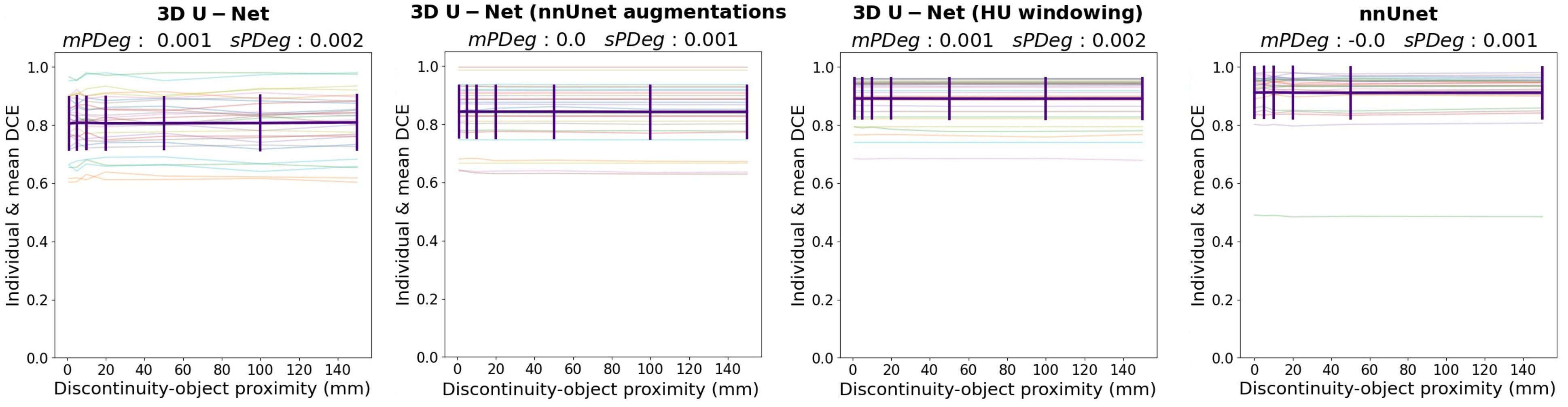}
	\caption{The dice (DSC) score produced by each tested model for each of the 30 test cases in the liver segmentation dataset, with simulated sudden patient motion by 10 degrees such that the resulting discontinuity at varying proximity to the liver ground truth segmentation. The corresponding means of the dice score across the test cases is shown in purple, with the error bars representing the standard deviation. The results in this plot are summarized in Tables \ref{tab_dce}.}
    \end{figure}

\end{document}